\documentclass[floats,floatfix,showpacs,amssymb,prd,twocolumn,superscriptaddress,nofootinbib,nolongbibliography,reprint]{revtex4-1}

\usepackage{amssymb,amsmath,verbatim,mathtools,needspace,enumitem,etoolbox,graphicx,physics,microtype,afterpage,xspace,tabularx,lmodern,multirow,bm}
\usepackage{dcolumn}
\usepackage{multirow}
\usepackage{float}
\usepackage{booktabs}
\usepackage{gensymb}
\usepackage{lipsum}
\usepackage[dvipsnames, usenames]{xcolor}
\definecolor{linkcolor}{rgb}{0.0,0.3,0.5}
\usepackage[unicode, colorlinks=true, linkcolor=linkcolor, citecolor=linkcolor, filecolor=linkcolor, urlcolor=linkcolor, linktocpage, breaklinks]{hyperref}
\usepackage[all]{hypcap}
\usepackage[T1]{fontenc}
\usepackage[utf8]{inputenc}
\usepackage[usenames,dvipsnames]{xcolor}
\hypersetup{colorlinks=true,citecolor=romared,linkcolor=romared,urlcolor=romared}

\definecolor{romared}{RGB}{142,0,28}

\newcommand{\be}{\begin{equation}}
\newcommand{\ee}{\end{equation}}

\def\be{\begin{equation}}
\def\ee{\end{equation}}
\newcommand{\beq}{\begin{eqnarray}}
\newcommand{\eeq}{\end{eqnarray}}

\usepackage{aas_macros}
\usepackage{makecell}
\usepackage{soul}
\usepackage[nolist,nohyperlinks]{acronym}

\usepackage [english]{babel}
\usepackage [autostyle, english = american]{csquotes}
\MakeOuterQuote{"}

\newcommand{\approptoinn}[2]{\mathrel{\vcenter{
  \offinterlineskip\halign{\hfil$##$\cr
    #1\propto\cr\noalign{\kern2pt}#1\sim\cr\noalign{\kern-2pt}}}}}

\newcommand{\lensMass}{M_{\rm Lz}}
\newcommand{\critLensMass}{\lensMass^{\rm crit}}
\newcommand{\critImpactParam}{y^{\rm crit}}
\newcommand{\Msun}{M_{\odot}}
\newcommand{\orderOf}{\mathcal{O}}

\begin{document}

\title{Observability of lensing of gravitational waves\\from massive black hole binaries with LISA}
\date{\today} 

\author{Mesut \c{C}al{\i}\c{s}kan}
\email{caliskan@jhu.edu}
\affiliation{William H. Miller III Department of Physics and Astronomy, Johns Hopkins University, Baltimore, MD 21218, USA}

\author{Lingyuan Ji}
\email{lingyuan.ji@jhu.edu}
\affiliation{William H. Miller III Department of Physics and Astronomy, Johns Hopkins University, Baltimore, MD 21218, USA}

\author{Roberto Cotesta}
\affiliation{William H. Miller III Department of Physics and Astronomy, Johns Hopkins University, Baltimore, MD 21218, USA}

\author{Emanuele Berti}
\affiliation{William H. Miller III Department of Physics and Astronomy, Johns Hopkins University, Baltimore, MD 21218, USA}

\author{Marc Kamionkowski}
\affiliation{William H. Miller III Department of Physics and Astronomy, Johns Hopkins University, Baltimore, MD 21218, USA}

\author{Sylvain Marsat}
\affiliation{Laboratoire des 2 Infinis - Toulouse (L2IT-IN2P3), Univers\'it\'e de Toulouse, CNRS, UPS, F-31062 Toulouse Cedex 9, France}

\begin{abstract}
\noindent
The gravitational waves emitted by massive black hole binaries in the LISA band can be lensed. 
Wave-optics effects in the lensed signal are crucial when the Schwarzschild radius of the lens is smaller than the wavelength of the radiation.
These frequency-dependent effects can enable us to infer the lens parameters, possibly with a single detection alone.
In this work, we assess the observability of wave-optics effects with LISA by performing an information-matrix analysis using analytical solutions for both point-mass and singular isothermal sphere lenses.
We use gravitational-waveform models that include the merger, ringdown, higher harmonics, and aligned spins to study how waveform models and source parameters affect the measurement errors in the lens parameters.
We find that previous work underestimated the observability of wave-optics effects and that LISA can detect lensed signals with higher impact parameters and lower lens masses.
\end{abstract}

\maketitle

\section{Introduction}

When electromagnetic (EM) waves travel near massive objects over cosmological distances, they get gravitationally lensed~\cite{Bartelmann:2010fz}.
Gravitational lensing leads to many exciting observations in the EM band, such as distortions of galaxy images into long arcs or ``Einstein rings,'' multiple images of the same supernova explosion, and statistical distortions of background radiation in the limit of weak lensing.
Gravitational lensing of EM waves is widely utilized in cosmology, astrophysics, and astronomy to reveal evidence of dark matter~\citep{Clowe:2003tk,Markevitch:2003at}, discover exoplanets~\citep{Bond:2004qd}, measure the Hubble constant~\citep{2016A&ARv..24...11T}, and uncover massive objects and structures that are too faint to be detected directly~\citep{Coe:2012wj}, for example. 

Just like EM waves, gravitational waves (GWs) can also get gravitationally lensed~\cite{Ohanian:1974ys,Thorne:1982cv,Deguchi:1986zz,Wang:1996as,Nakamura:1997sw,Takahashi:2003ix}. 
If observed, lensed GWs could enable a plethora of new scientific studies. 
When combined with EM lensing surveys, they may allow us to locate merging black holes at a sub-arcsecond precision~\cite{Hannuksela:2020xor}.
If accompanied by an EM counterpart, the sub-millisecond lensing time-delay measurements granted by GW observations could enable precision cosmography~\cite{Sereno:2011ty, Liao:2017ioi, Cao:2019kgn, Li:2019rns, Hannuksela:2020xor, Yu:2020agu}.
It has also been suggested that lensed GWs can be used to measure the speed and polarization content of GWs~\cite{Baker:2016reh, Fan:2016swi,Goyal:2020bkm}, detect intermediate-mass and primordial black holes through micro-lensing~\cite{Lai:2018rto, Diego:2019rzc, Oguri:2020ldf}, and constrain the population of lenses~\cite{Xu:2021bfn}.

The prospect of observing GW lensing at low frequencies with the Laser Interferometer Space Antenna (LISA)~\cite{2017arXiv170200786A} is particularly exciting. 
While the geometric-optics approximation holds for the strongly-lensed stellar-mass black-hole binary (BHB) mergers accessible to the ground-based GW detectors such as LIGO~\cite{LIGOScientific:2014pky}, Virgo~\cite{VIRGO:2014yos}, and KAGRA~\cite{KAGRA:2020cvd, Ohanian:1974ys, Caliskan:2022wbh, Dai:2020tpj, Deguchi:1986zz, Takahashi:2003ix, Hannuksela:2019kle}, the massive black-hole binaries (MBHBs) detectable by LISA emit GWs at much lower frequencies, allowing the possibility for wave-optics effects (such as diffraction) to be detected in the lensed signal. If the Schwarzschild radius of the lens is smaller than the wavelength $\lambda$ of the GWs, diffraction effects are crucial.
For diffraction to be prominent, the lens mass $M_{\rm L}$ must satisfy the condition~\cite{Takahashi:2003ix}
\begin{equation}\label{eqn:diffraction_condition}
    M_{\rm L} \lesssim 10^5\, M_\odot \left(\frac{f}{\textrm{Hz}}
    \right)^{-1},
\end{equation}
where $f$ is the GW frequency.

Wave-optics effects can lead to frequency-dependent amplitude and phase modulations in the GW detections.
Therefore, LISA detections of these lensing-induced effects may be used to measure the lens parameters, such as the redshifted lens mass $\lensMass = (1+z_{\rm L})M_{\rm L}$, where $z_{\rm L}$ is the redshift of the lens, and the position of the source in the source plane. More ambitiously, if the event rates are large enough, the measurement of lens parameters may even enable us to probe the lens population. 
Furthermore, the characteristic interference patterns observed in the signal can be used to break the so-called mass-sheet degeneracy, in part of the wave-optics regime and in the interference regime, with only one lensed waveform~\cite{Cremonese:2021puh}.

Wave-optics effects in gravitational lensing of GWs have been extensively studied in the literature~\cite{Takahashi:2003ix, Cremonese:2021puh, Gao:2021sxw, Ohanian:1974ys, Nakamura:1997sw, DePaolis:2002tw, Oguri:2020ldf, Takahashi:2004mc, Takahashi:2005ug, Meena:2019ate}.
In their pioneering work, Takahashi and Nakamura~\cite{Takahashi:2003ix} (henceforth TN) calculated how accurately the lens parameters could be measured using an information-matrix analysis. They considered GWs lensed by either point-mass (PM) or singular isothermal sphere (SIS) lenses in the mass range $\lensMass \in [10^6,\ 10^9] \, \Msun$.
For a LISA MBHB with detector-frame (redshifted) total mass $M_{\rm Tz} = 2 \times 10^6 \, \Msun$, and mass ratio $q=1$, they found that wave-optics effects allow for the measurement of the lens parameters for SIS lenses in the range $\lensMass \approx 10^6 - 10^8 \, \Msun$.
However, TN found that lensing magnification is negligible and that the lens parameters are not well measured for $\lensMass \lesssim 10^6 \, \Msun$; therefore, they did not investigate the case of lower lens masses.

Recent work~\cite{Gao:2021sxw} claimed that over $(0.1-1.6)\%$ of the MBHBs with total (source-frame) mass $10^{5} - 10^{6.5} \, \Msun$ and redshift $z_{\rm S}= 4-10$ could have wave-optics effects detectable by LISA even when the impact parameter $y$ is as large as $y \simeq 50$.
This claim is noteworthy for three reasons: 
(i) if robust, the lensing probability could be an order of magnitude larger than what was claimed in previous work; 
(ii) TN found that, for SIS lenses, wave-optics effects would be detectable for impact parameters as high as $y \sim 3$, considerably smaller than the value ($y \simeq 50$) found in~\cite{Gao:2021sxw}; 
and, (iii) according to Ref.~\cite{Gao:2021sxw}, wave-optics effects could be distinguishable for SIS lenses with $\lensMass = 10^1 - 10^4\, \Msun$, several orders of magnitude smaller than the value of $\lensMass=10^6\, \Msun$ considered in the TN study. These interesting claims motivated us to revisit the problem.

The authors of Ref.~\cite{Gao:2021sxw} defined detectability in terms of the so-called ``Lindblom criterion'' ~\cite{Flanagan:1997kp,Lindblom:2008cm,McWilliams:2010eq,Chatziioannou:2017tdw}, i.e., they assumed
the difference $\delta h \equiv h_{\rm L} - h_{\rm U}$ between the lensed waveform $h_{\rm L}$ and the unlensed waveform $h_{\rm U}$ to be discernible when $\langle \delta h | \delta h \rangle > 1$.
According to this rough criterion, the wave-optics effects are measurable if the signal-to-noise ratio (SNR) of the difference between the lensed and unlensed waveform is greater than 1. 
The criterion may be too optimistic because it assumes that the deviations from the theoretical waveform are solely due to lensing and might not account for possible degeneracies between the source and lens parameters (see, e.g.,~\cite{Ezquiaga:2020gdt}). 

One of the main goals of this paper is to update the pioneering TN exploration of the detectability and measurability of lensing effects in the GW signals emitted by MBHBs. 
The TN study predated the 2005 numerical relativity breakthrough, and, therefore, used an inspiral-only waveform based on the restricted post-Newtonian approximation, which does not take into account the merger, ringdown, and higher-order modes. 
In this work we use two waveform models: (i) \texttt{IMRPhenomD}, a (quadrupole-only) phenomenological waveform model describing the full inspiral, merger, and ringdown of aligned-spin BHBs~\cite{Husa:2015iqa,Khan:2015jqa}, and (ii) \texttt{IMRPhenomHM}, a phenomenological waveform model that also includes the higher-order modes~\cite{London:2017bcn}. 
The comparison between \texttt{IMRPhenomD} and \texttt{IMRPhenomHM} allows us to investigate the effects of higher-order modes on the measurability of lensing.

In their study, TN approximated lensed waveforms using either the geometric-optics limit or the short-time-delay limit. 
They also used the low-frequency approximation for the detector response, as opposed to the full response.
We use analytical solutions to the lensing diffraction integral in the wave-optics regime for both PM and SIS lenses and use these solutions to obtain analytical derivatives of the lensing diffraction integral.
We use these analytical derivatives to determine the precision with which the lens parameters can be measured by extending the information-matrix calculation implemented in the \texttt{lisabeta} code~\cite{Marsat:2018oam}, which computes the LISA detector response in the Fourier domain.
Our 13-dimensional matrices include all source parameters (including aligned spins) as well as the lens parameters and account for possible degeneracies between them.
In this way, we can estimate the errors in the lens parameters for MBHBs in a wide range of lens masses $\lensMass \in [10^{1}, 10^9]\, \Msun$ and impact parameters $y \in [0.01, 200]$.

The paper is organized as follows. In Sec.~\ref{sec:lensing_and_WO}, we review wave-optics effects in the gravitational lensing of GWs and provide analytical solutions to the diffraction integral for both PM and SIS lenses.
In Sec.~\ref{sec:LensedWaveforms}, we describe the effect of lensing on GWs and provide examples of lensed waveforms and the information-matrix formalism used to estimate measurement uncertainties in the MBHB and lens parameters.
In Sec.~\ref{sec:Results}, we discuss the measurement errors of lensing parameters, and in Sec.~\ref{sec:Conclusions}, we present conclusions and possible directions for future work. 
Throughout the paper, we assume a $\Lambda$CDM cosmology with cosmological parameter values matching Planck 2018~\cite{Planck:2018vyg}: Hubble constant $H_0 = 67.4\ \rm km\, s^{-1}\, Mpc^{-1}$, and
matter density $\Omega_{m} = 0.315$.
Unless specified otherwise, we work in geometrical units $(G=c=1)$.

\begin{figure*}[t!]
    \centering
    \includegraphics[width=\textwidth]{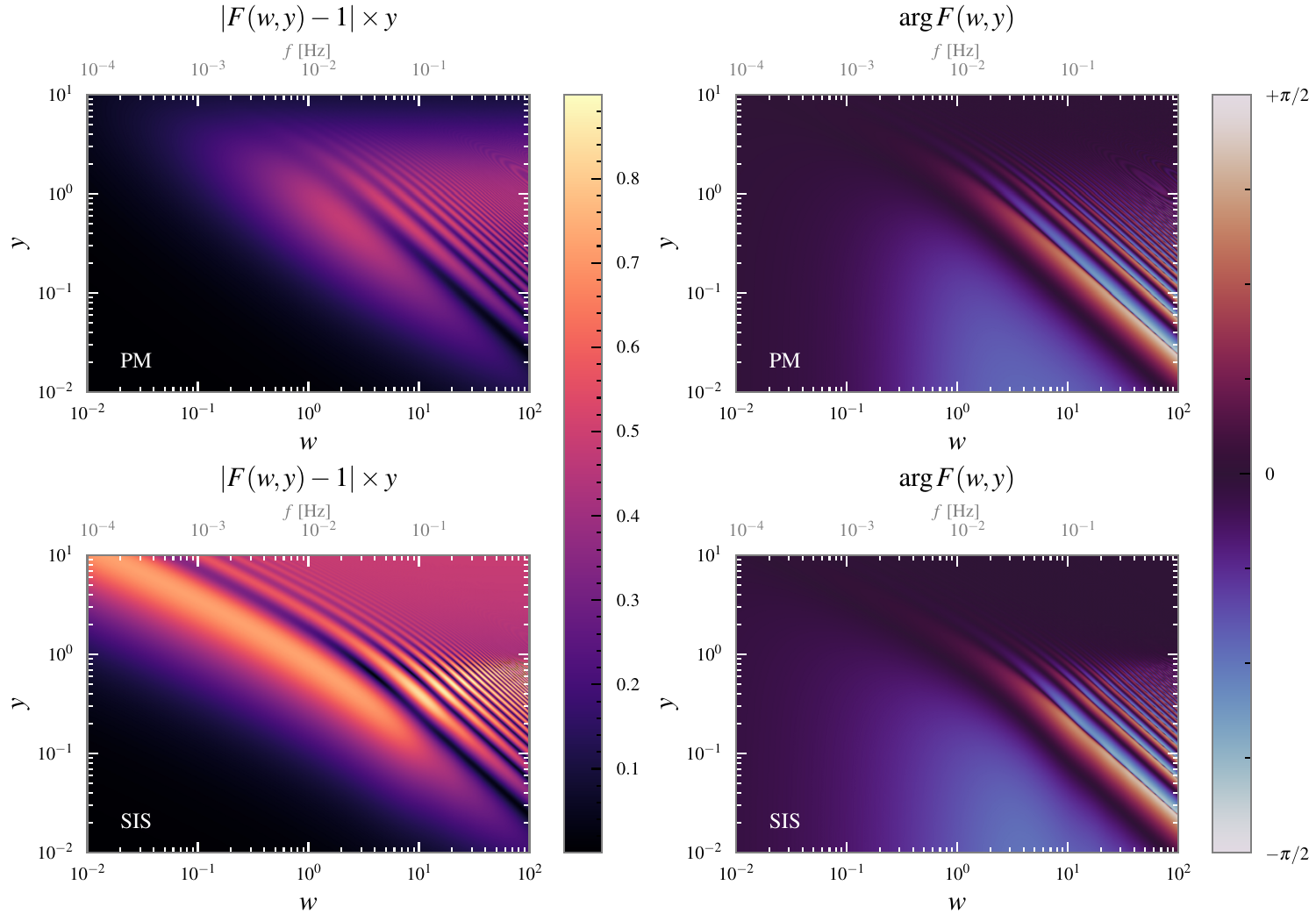}
    \caption{Left panels: absolute value of the diffraction integral ``contrast'' $|F(w,y) - 1|$, multiplied by the impact parameter $y$ to compensate for the dynamic range. Right panels: phase factor $\arg F(w,y)$. These quantities were computed by evaluating the diffraction integral for the PM lens (top row) and SIS lens (bottom row), and they are shown as functions of the dimensionless frequency $w = 8\pi M_{\rm Lz}f$ and impact parameter $y$. 
    The top x-axis in each panel shows the physical frequency (in Hz) corresponding to a redshifted lens mass $M_{\rm Lz} = 10^6\, \Msun$.
    }
    \label{fig:diffraction_integral}
\end{figure*}

\section{Gravitational Lensing and Wave Optics} \label{sec:lensing_and_WO}

The effect of a lens on GW propagation can be obtained by solving the complex-valued diffraction integral\footnote{We prefer to call this quantity the diffraction integral (rather than the ``amplification factor'', as it is also known) because, in the regime of interest for this paper, lensing can induce frequency-dependent modulations in both the amplitude and phase of the GWs.} for a given source frequency $f$~\cite{Takahashi:2003ix}:
\begin{equation}\label{eqn:amplification-factor}
    F(f, \bm{y}) = \frac{D_{\rm S} (1+z_{\rm L}) \xi_0^2}{D_{\rm L} D_{\rm LS}} \frac{f}{i} \int \mathrm{d}^2 \bm{x}\, \exp[2\pi i f t_{\rm d}(\bm{x}, \bm{y})].
\end{equation}
The integral is over all possible paths, including those which are not geodesics. Here, $D_{\rm L}$, $D_{\rm S}$, and $D_{\rm LS}$ are the angular-diameter distances from the observer to the lens, from the observer to the source, and from the lens to the source, respectively. 
The dimensionless 2-vectors $\bm{x}$ and $\bm{y}$ are defined as 
\begin{align}\label{eqn:impact_parameter}
    \bm{x} \equiv \frac{\bm{\xi}}{\xi_0} && \textrm{and} && \bm{y} \equiv \bm{\eta} \frac{D_{\rm L}}{\xi_0 D_{\rm S}},
\end{align}
where $\bm{\xi}$ and $\bm{\eta}$ are the physical coordinates of the image in the lens plane and of the source in the source plane, respectively.

The arbitrary length normalization $\xi_0$ is usually chosen to be the relevant scale of the problem. The time delay for a given path is defined as
\begin{equation}\label{eqn:time-delay}
    t_{\rm d}(\bm{x}, \bm{y}) = \frac{D_{\rm S} \xi_0^2}{D_{\rm L} D_{\rm LS}} (1+z_{\rm L})\left[\frac12 |\bm{x}-\bm{y}|^2 - \psi(\bm{x}) + \phi(\bm{y}) \right],
\end{equation}
where $\psi(\bm{x})$ is the deflection potential.
The quantity $\phi(\bm{y})$ sets the zero point of the time delay for a given source position $\bm{y}$, and it does not affect the relative time delay between different paths. 
For convenience, we set $\phi(\bm{y})$ so that the minimum possible time delay $\min_{\bm{x}} t_{\rm d}(\bm{x}, \bm{y})$ is zero.

From now on, for simplicity, we will restrict our discussion to spherically symmetric lenses. In this case, the problem becomes one-dimensional, so $\psi(\bm{x}) = \psi(x)$, and $\phi(\bm{y}) = \phi(y)$, where $x\equiv |\bm{x}|$ and $y\equiv |\bm{y}|$. 
Without loss of generality, the angular integral in Eq.~\eqref{eqn:amplification-factor} can be performed by aligning the reference direction of the polar coordinates with $\bm{y}$, resulting in
\begin{multline}\label{eqn:amplification-factor-spherical}
    F(w, y) = \frac{w}{i} \exp\left\{i w\left[\frac{y^2}{2}+\phi(y)\right]\right\} \\
    \times \int_0^\infty x\mathrm{d}x\, \exp\left\{i w\left[\frac{x^2}{2}-\psi(x)\right]\right\} J_0(w x y).
\end{multline}
Here,
\begin{equation}\label{eqn:w}
w \equiv \frac{D_{\rm S} \xi_0^2 (1+z_{\rm L}) (2\pi f)}{D_{\rm L} D_{\rm LS}}
\end{equation}
is a dimensionless frequency, and $J_0$ denotes the zeroth-order Bessel function. 
We will now apply Eq.~\eqref{eqn:amplification-factor-spherical} to two specific mass distributions.

\subsection{Point-mass lens}

Let us first consider the simple case of a PM lens, for which the mass density $\rho_\mathrm{PM}(\bm{r}) = M_{\mathrm L} \delta^3(\bm{r})$.
Here, $\delta^3(\bm{r})$ is the three-dimensional Dirac delta function.
A natural choice for $\xi_0$ is the Einstein radius, i.e.,
\begin{equation}
\xi_0 =\left (\frac{4 M_{\rm L} D_{\rm L} D_{\rm LS}}{D_{\rm S}}\right)^{1/2}. 
\end{equation}
With this choice, we have $\psi(x) = \ln x$, and the radial integral can be solved analytically with the result~\cite{Takahashi:2003ix}
\begin{multline}\label{eqn:F_point_mass}
    F(w,y) = \exp\left\{\frac{\pi w}{4} + i\frac{w}{2}\left[\ln\frac{w}{2} - 2\phi(y)\right]\right\} \\ \times \Gamma\left(1-\frac{w}{2}i\right) {_1 F_1} \left(\frac{w}{2}i, 1; \frac{wy^2}{2}i \right).
\end{multline}
Here, $w = 8\pi M_{\rm L} (1+z_{\rm L}) f$, $\phi(y) = (x_+ - y)^2/2 - \ln x_+$, $x_+ = [(y^2+4)^{1/2}+y]/2$, and ${_1 F_1}(a, b; z)$ is the confluent hypergeometric function.

\subsection{Singular isothermal sphere lens} \label{sec:SIS_lens}

For a singular isothermal sphere with velocity dispersion $\sigma_v$, the mass density reads $\rho_\mathrm{SIS}(\bm{r}) = \sigma_v^2 / (2\pi |\bm{r}|^2)$.
For an axially-symmetric gravitational lens, the lens mass $M_{\rm L}$ is defined as the amount of mass enclosed within the Einstein radius of the lens. 
Therefore, the total mass of the lens and the lens mass are equivalent for a point-mass lens. 
This may not be the case for other lens profiles; for example, the total mass of a dark matter halo with the SIS profile is different from the lens mass of the halo. 
For the SIS profile, the lens mass is related to $\sigma_v$ as
\begin{equation}
  M_{\rm L} = \frac{4 \pi^2 \sigma_v^4 D_{\rm L}D_{\rm LS}}{D_{\rm S}}.
  \label{eq:MLsigmav}
\end{equation}
The Einstein radius is 
\begin{equation}
    \xi_0 = \frac{4\pi \sigma_v^2 D_{\rm L} D_{\rm LS}}{D_{\rm S}} = \left (\frac{4 M_{\rm L} D_{\rm L} D_{\rm LS}}{D_{\rm S}}\right)^{1/2},
\end{equation}
and $w = 8\pi M_{\rm L}(1+z_{\rm L})f$ (as in the case of a PM lens).

We choose the normalization to be $\xi_0$, giving $\psi(x) = x$ and $\phi(y)=y+1/2$ in Eq.~\eqref{eqn:amplification-factor-spherical} for the SIS lens.
The resulting formula for the radial integral can be found in Ref.~\cite{Takahashi:2003ix}.
The numerical evaluation of this formula is difficult for large values of $w$ and $y$ since both the exponential and Bessel-function factors in the integrand can oscillate rapidly.

Several different numerical approaches have been proposed to tackle this problem~\cite{Ulmer:1994ij,1999PThPS.133..137N,Takahashi:2003ix}. 
Here, we propose and implement a simple, effective method based on a Taylor expansion\footnote{After our pre-print appeared on the arXiv, Ryuichi Takahashi brought to our attention that a perturbative expansion of the lensing potential to find analytical solution of the diffraction integral for SIS lenses was also proposed in Ref.~\cite{Matsunaga:2006uc}.} of the exponential factor $\exp[-i w \psi(x)]$ in Eq.~\eqref{eqn:amplification-factor-spherical}. 
We begin our evaluation by defining the integral\footnote{This integral corresponds to Eq.~(6.631.1) in Ref.~\cite{gradshteyn2007} if we make the substitutions $\alpha = -iw/2$, $\beta = wy$, $\mu = n+1$, and $\nu = 0$.}
\begin{align}
    I_n(w,y) & \equiv \int_0^\infty x^n e^{i w x^2/2} J_0(wxy)\, x\mathrm{d}x \\
            \label{eqn:important_integral}
             & = \frac12 \left(\frac{2i}{w}\right)^N \Gamma\left(N\right) {_1 F_1} \left(N, 1; -i\frac{wy^2}{2}\right),
\end{align}
where $N \equiv (n+2)/2$.
We also define the series expansion of the exponential of the potential, 
\begin{equation}
    \Psi(w, x) \equiv e^{-i w \psi(x)} = \sum_{n=0}^\infty \Psi_n(w) x^n.
\end{equation}
Using these definitions, Eq.~\eqref{eqn:amplification-factor-spherical} can then be evaluated by integrating the expansion term by term, which gives
\begin{equation}\label{eqn:amplification-factor-spherical-series}
    F(w, y) = \frac{w}{i} \exp\left\{i w\left[\frac{y^2}{2}+\phi(y)\right]\right\} \sum_{n=0}^\infty \Psi_n(w) I_n (w, y).
\end{equation}
This series is usually only conditionally convergent or even divergent, but it can be summed with the help of series acceleration techniques.
For an SIS lens, $\psi(x) = x$, so $\Psi_n(w) = (-iw)^n/n!$, and the Shanks transformation (e.g., see~\cite{Press:1992zz}) performs well in accelerating the summation. When $(w, y)$ approaches the geometric-optics limit, this series requires very-high floating-point precision and sufficiently many terms to provide satisfactory convergence. 
So, in practice, we use a piecewise strategy to evaluate $F(w, y)$ with the help of geometric-optics approximation, detailed in Appendix~\ref{sec:SIS-evaluation}. 

Diffraction integrals computed using Eq.~\eqref{eqn:F_point_mass} (for PM lenses, top panels) and this analytical solution (for SIS lenses, bottom panels) are shown in Fig.~\ref{fig:diffraction_integral}.
For a given impact parameter $y$, at sufficiently small values of the dimensionless frequency $w$, the lensing effect is negligible because the lens size is negligible compared to the wavelength of GWs. 
As $w$ increases, the effect of lensing starts to be visible through the oscillations of $F(w, y)$ as a function of $w$. The value of $w$ marking the transition between these two regimes depends on the impact parameter $y$.

In closing this section, let us note that the total mass of the SIS profile is, strictly speaking, infinite. This nonphysical behavior is conventionally regularized by introducing an outer boundary at $r = r_\Delta$ such that $\rho_\mathrm{SIS}(r_\Delta) = \Delta \rho_\mathrm{cr}$, where $\Delta$ is a dimensionless constant (we set $\Delta = 200$), and $\rho_\mathrm{cr} = 3H_\mathrm{L}^2/(8\pi)$ is the critical density of the Universe at the redshift $z_\mathrm{L}$ with the corresponding Hubble parameter $H_\mathrm{L}$. In Appendix~\ref{sec:SIS-boundary}, we demonstrate that this truncation does not affect our results.

\begin{figure*}[t!]
    \centering
    \includegraphics[width=\textwidth]{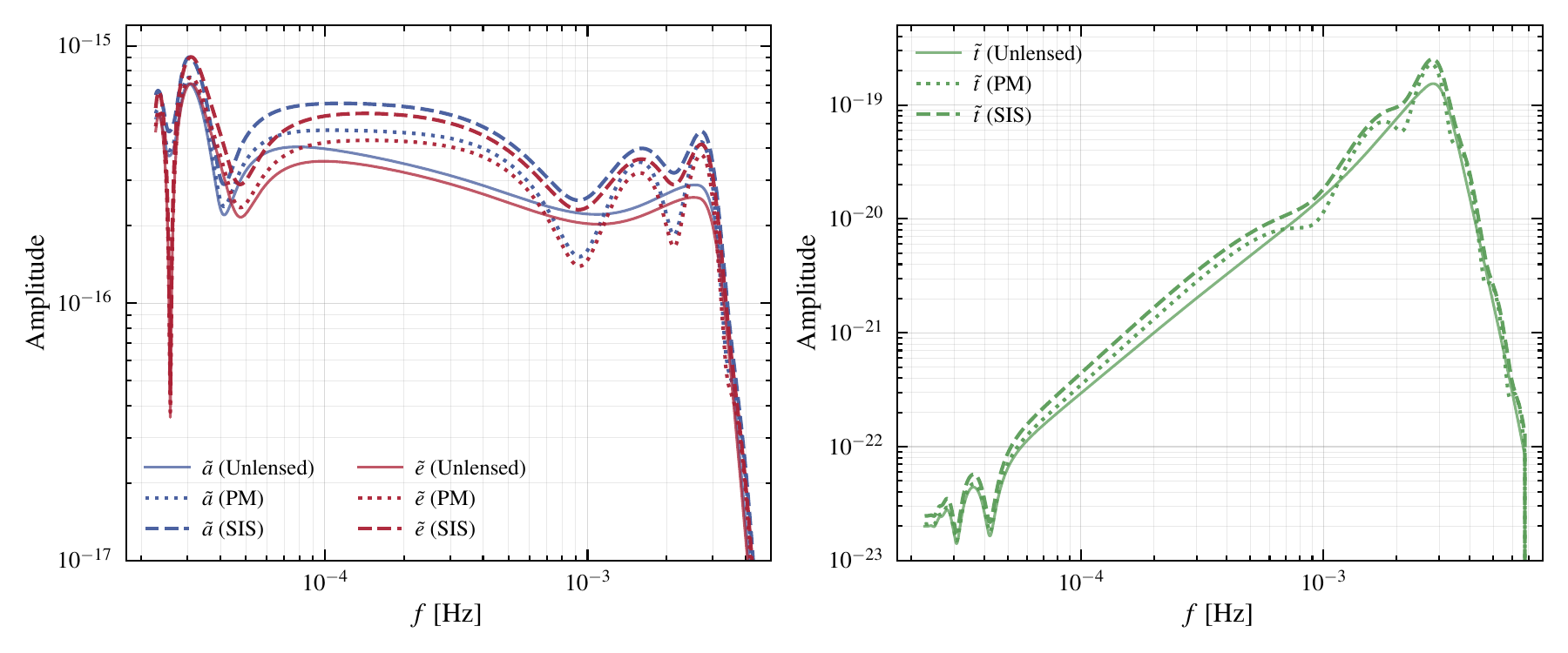}
    \caption{Comparison between the frequency-domain amplitude of unlensed waveforms (solid lines) and waveforms lensed by a PM (dotted) or SIS (dashed) lens. In the left panel, blue and red lines refer to the TDI observables $\tilde{a}$ and $\tilde{e}$, respectively. The right panel shows the TDI observable $\tilde{t}$, where the signal amplitude is much smaller. All results refer to a redshifted lens mass $M_{\rm Lz} = 2\times 10^7\ \rm M_{\odot}$, a lens redshift $z_{\rm L} = 1$, and an impact parameter $y=1.0$. The source parameters are $M_{\rm Tz} = 6 \times 10^6\ M_{\odot}$, $z_{\rm S} = 2$, $\iota = 2.42$, $\phi_c = 1.84$, $\lambda = 0.3$, $\beta = 0.3$, $\psi=0.94$, and $\chi_m = \chi_p = 0$.
    }
    \label{fig:lensed_example}
\end{figure*}

\section{Lensed gravitational waveforms and information-matrix formalism} \label{sec:LensedWaveforms}

The lensed gravitational waveform in the frequency domain $\Tilde{h}^{\rm L}(f;\bm{\theta}^{\rm S}) \equiv \Tilde{h}_+^{\rm L} - i\Tilde{h}_\times^{\rm L}$ is given by the product of the diffraction integral $F(w,y)$ and the unlensed waveform $\Tilde{h}(f)$
\begin{equation}\label{eqn:lensed_strain}
    \Tilde{h}^{\rm L}(f;\bm{\theta}^{\rm S}, \bm{\theta}^{\rm L}) = F(w,y) \Tilde{h}(f;\bm{\theta}^{\rm S})\,,
\end{equation}
where $w=8\pi M_{\rm Lz} f$, and $y$ and $F(w,y)$ are given by Eqs.~\eqref{eqn:impact_parameter} and~\eqref{eqn:amplification-factor-spherical}, respectively. The vector $\bm{\theta}^{\rm S} \equiv \{M_{\rm Tz},\ q,\ d_{l},\ t_{\rm c},\ \iota,\ \phi_{\rm c},\ \lambda,\ \beta,\ \psi,\ \chi_{\rm m},\ \chi_{\rm p} \}$ includes 11 source parameters: the detector-frame total mass $M_{\rm Tz}$, mass ratio $q$, luminosity distance to the source $d_{l}$, coalescence time $t_{\rm c}$, inclination angle $\iota$, coalescence phase $\phi_{\rm c}$, right ascension $\lambda$, declination $\beta$, polarization angle $\psi$, and two parameters -- the ``effective spin'' $\chi_{\rm p}=(m_1\chi_1+m_2\chi_2)/(m_1+m_2)$
and the asymmetric spin combination $\chi_{\rm m}=(m_1\chi_1-m_2\chi_2)/(m_1+m_2)$ --
for the spins of the binary components, which we assume to be aligned with the orbital angular momentum. 
The vector $\bm{\theta}^{\rm L} \equiv \{M_{\rm Lz},\ y\}$ includes, in contrast, the lens parameters.
Using the decomposition of the waveform in spin-weighted spherical harmonics $\Tilde{h}(f;\bm{\theta}^{\rm S}) = \sum_{\ell, m} {}_{-2}Y_{\ell m}\Tilde{h}_{\ell m}(f;\bm{\theta}^{\rm S})$, and Eq.~\eqref{eqn:lensed_strain}, it is straightforward to derive the expression of the lensed GW modes $\Tilde{h}_{\ell m}^{\rm L}(f;\bm{\theta}^{\rm S})$ as
\begin{equation}
\label{eq:lensed_modes}
\Tilde{h}_{\ell m}^{\rm L}(f;\bm{\theta}^{\rm S}, \bm{\theta}^{\rm L}) = F(w,y)\Tilde{h}_{\ell m}(f;\bm{\theta}^{\rm S}).
\end{equation}

A GW signal causes a shift in the frequency of the laser traveling between spacecraft pairs in the LISA constellation. 
This effect can be described using three reduced time-delay interferometry (TDI) observables: $\Tilde{a}(f)$, $\Tilde{e}(f)$ and $\Tilde{t}(f)$. 
These observables are mutually independent, and they represent a particular combination of the shifts in the laser frequency between spacecraft pairs that reduces the effect of the laser noise. 
Their definition is
\begin{equation}
\label{eq:reduced_TDI}
    \Tilde{a}(f), \Tilde{e}(f), \Tilde{t}(f) = \sum_{\ell m} T_{a,e,t}^{\ell m}(f) \Tilde{h}_{\ell m}(f;\bm{\theta}^{\rm S}),
\end{equation}
where $T_{a,e,t}^{\ell m}(f)$ are mode-by-mode transfer functions describing the response of the LISA detector to the passage of the GW signal defined in Eq.~(20) of Ref.~\cite{Marsat:2020rtl}, and $\Tilde{h}_{\ell m}(f)$ are the modes of the gravitational radiation crossing the LISA detector. 
In the case of a lensed signal, the GW modes $\Tilde{h}_{\ell m}(f;\bm{\theta}^{\rm S})$ should be replaced by the lensed modes $\Tilde{h}_{\ell m}^{\rm L}(f;\bm{\theta}^{\rm S}, \bm{\theta}^{\rm L})$ defined in Eq.~\eqref{eq:lensed_modes}.

In Fig.~\ref{fig:lensed_example}, we show the amplitude of the three reduced TDI observables as a function of the GW frequency for a reference source with binary parameters $M_{\rm Tz} = 6 \times 10^6\ M_{\odot}$, $z_{\rm S} = 2$, $\iota = 2.42$, $\phi_c = 1.84$, $\lambda = 0.3$, $\beta = 0.3$, and $\psi=0.94$. 
We compare the unlensed GW signal with the same signals lensed by either PM or SIS lenses with redshift $z_{\rm L} = 1$, redshifted lens mass $M_{\rm Lz} = 2\times 10^7\ \rm M_{\odot}$, and impact parameter $y=1.0$. 
Diffraction effects are clearly visible, and the amplitude and frequency of the wave-optics modulations depend on the structure of the lens. 
Both the PM and the SIS lenses induce strong oscillations in the amplitude of the TDI reduced observables at frequencies around $0.1$~mHz because lensing causes GWs to travel through different path lengths and therefore produces interference.

We use the information-matrix formalism (or linear signal approximation)~\cite{Cutler:1997ta,Takahashi:2003ix,Berti:2004bd,Cutler:2007mi} to determine the uncertainties in estimating the parameters of the MBHB system and the lens. 
This formalism is valid in the large-SNR limit, and therefore it is expected to be accurate for most LISA MBHBs. 
In the linear signal approximation, the likelihood associated with each reduced TDI observable in Eq.~\eqref{eq:reduced_TDI} is a multidimensional Gaussian of the form
\begin{equation}
    p(\Delta \theta_i) = \mathcal{N} \exp{-\frac{1}{2}\Gamma_{ij}^{\Tilde{X}}\Delta \theta_i \Delta \theta_j}\,,
\end{equation}
where $\bm{\theta} \equiv \{\bm{\theta}^{\rm S},\bm{\theta}^{\rm L}\}$, $\Gamma_{ij}^{\Tilde{X}}$ is the information matrix associated to each observable $\Tilde{X} \in \{ \Tilde{a}, \Tilde{e}, \Tilde{t} \}$, and $\mathcal{N} = \sqrt{\det(\Gamma/2\pi)}$ is a normalization factor. The information matrix for each reduced TDI observable reads
\begin{equation}\label{eq:TDI_fisher}
   \Gamma_{ij}^{\Tilde{X}} = \left ( \frac{\partial\Tilde{X}}{\partial \theta_{i}} \Bigg | \frac{\partial\Tilde{X}}{\partial \theta_{j}} \right ),
\end{equation}
where the inner product is defined as
\begin{equation}\label{eqn:inner_product}
    (a|b) \equiv 4 \Re \int_{0}^{\infty} \textrm{d}f \frac{\tilde{a}(f)\tilde{b}^*(f)}{S_n(f)},
\end{equation}
and $S_n(f)$ is the \texttt{SciRDv1}~\cite{SciRDv1} LISA power spectral density (PSD).
In practice, for each binary, we fixed the initial frequency of the integral in Eq.~\eqref{eqn:inner_product} to obtain a time-to-merger of at most one year, with a lower boundary of $f_{\rm min} = 10^{-5}\ \textrm{Hz}$.
The detailed calculation of the derivatives of $\Tilde{X}$ appearing in Eq.~\eqref{eq:TDI_fisher} is given in Appendix~\ref{app:analytic_derivatives}. 
Since the TDI observables are independent, the total likelihood is the product of the likelihoods, and therefore the total information matrix is 
\begin{equation} \label{eqn:information_matrix_summation}
    \Gamma = \Gamma_{ij}^{\Tilde{a}}+ \Gamma_{ij}^{\Tilde{e}} + \Gamma_{ij}^{\Tilde{t}}\,.
\end{equation}
The uncertainties on the parameters $\bm{\theta}$ can then be found from the variance-covariance matrix (the inverse of the information matrix):
\begin{equation}\label{eqn:variance-covariance}
  \langle \Delta\theta^{i} \Delta\theta^{j} \rangle = (\Gamma^{-1})^{ij}\,.
\end{equation}

\begin{figure*}[t!]
    \centering
    \includegraphics[width=\textwidth]{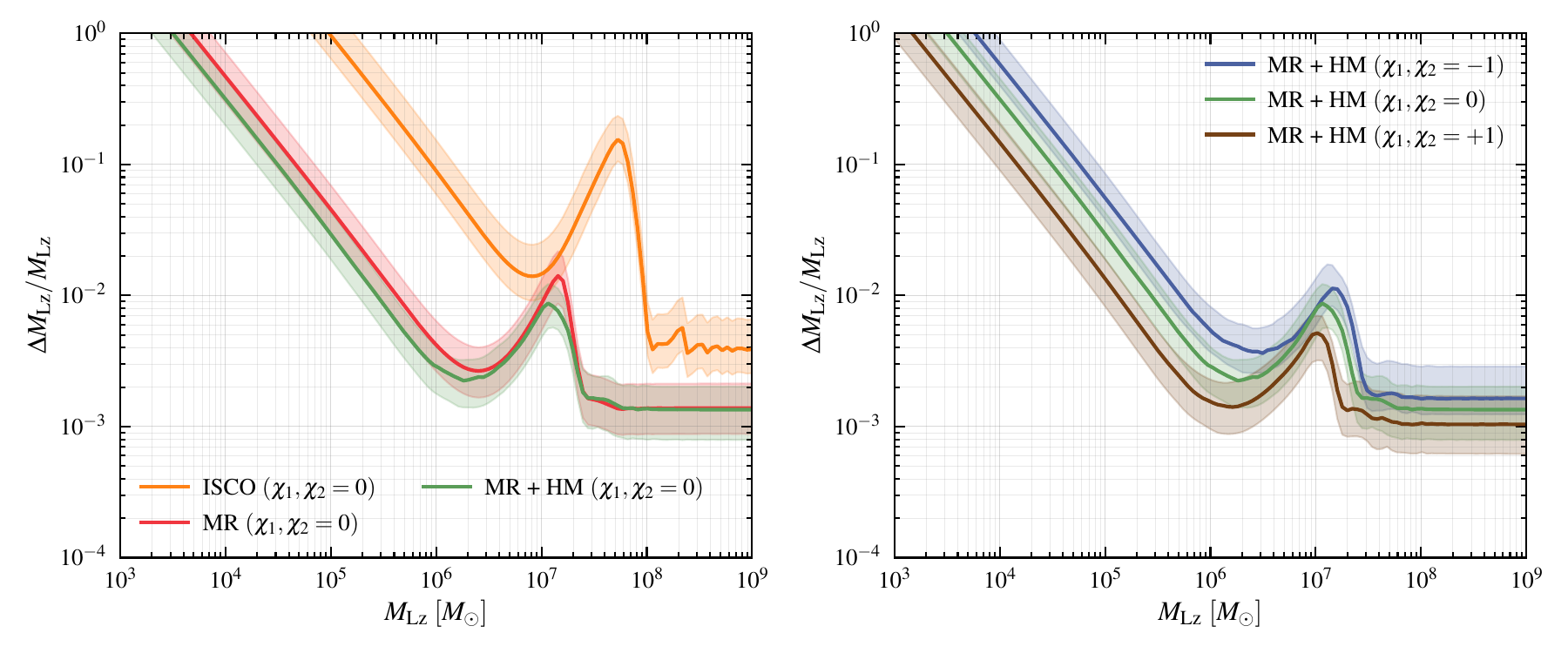}
    \caption{Relative uncertainty on the redshifted lens mass, $\Delta \lensMass/\lensMass$. Left: comparison between the ``inspiral only'' version of \texttt{IMRPhenomD} truncated at the ISCO (orange), the \texttt{IMRPhenomD} model including the merger and ringdown (red), and the \texttt{IMRPhenomHM} including also the higher harmonics (green). Right: comparison of \texttt{IMRPhenomHM} waveforms with different spin magnitudes. The green curve refers to nonspinning binaries, while the blue (brown) curves refer to the \texttt{IMRPhenomHM-} (\texttt{IMRPhenomHM+}) extremal anti-aligned (aligned) spin models.
    All results are for $\orderOf(100)$ MBHBs with $M_{\rm Tz} = 2 \times 10^6\, \Msun$, $q=1$, and $z_{\rm S} = 1$, with extrinsic parameters ($\iota$, $\phi_c$, $\lambda$, $\beta$, and $\phi$) randomly sampled over uniform distributions.
    Dark solid lines show the median value of $\Delta \lensMass/\lensMass$, while the shaded regions correspond to $1\sigma$ confidence intervals.
    Here we consider PM lenses with a range of redshifted lens masses $\lensMass$, but we fix the impact parameter to $y=0.1$.
  }%
    \label{fig:waveformComparison}%
\end{figure*}

\section{Results} \label{sec:Results}

In this section, we investigate how waveform models and source parameters affect the measurement of the lens parameters for both PM and SIS lenses. 
We first investigate how the measurement of the lens parameters is affected by different waveform models (Sec.~\ref{sec:Results_HOmodes}) and source parameters (Sec.~\ref{sec:EffectsOfTheSourceParameters}).
Then we present an extensive exploration of lens-parameter-measurement accuracy for a wide range of lens masses and impact parameters, considering first PM lenses
(Sec.~\ref{sec:Results_PML}) and then SIS lenses (Sec.~\ref{sec:Results_SIS}).

\subsection{Effect of the merger, ringdown, and higher-order modes on the measurement of the lens parameters} \label{sec:Results_HOmodes}

Our goal is to improve over the pioneering TN analysis~\cite{Takahashi:2003ix} by considering the effects of the merger, ringdown, and higher harmonics. 
In their work, the errors in the lens mass and impact parameter were estimated for a single MBHB and scaled by the source SNR to estimate the measurement uncertainty of lens parameters for other lensed MBHBs. 
More importantly, the TN analysis predated the 2005 numerical relativity breakthrough and therefore neither included the merger and ringdown nor higher harmonics. 
Furthermore, their work does not include an exploration of how other source parameters (such as the binary's inclination angle and component spins) affect the measurement of the lens parameters. 

We estimate measurement uncertainties on $\lensMass$ and $y$ for $\orderOf(100)$ MBHBs with fixed intrinsic parameters (to begin with, we fix the detector-frame total mass $M_{\rm Tz} = 2 \times 10^6\, \Msun$, mass ratio $q=1$, and redshift $z_{\rm S} = 1$), and randomly sampled extrinsic parameters (inclination angle $\iota$, coalescence phase $\phi_c$, right ascension $\lambda$, declination $\beta$, and polarization angle $\psi$) over uniform distributions.
To understand the effect of the merger/ringdown, higher-order modes, and spins, we focus on four representative waveform models: (i) a nonspinning \texttt{IMRPhenomD} model where the signal is truncated at the innermost stable circular orbit (henceforth ISCO), which includes only the inspiral part of the waveform and closely mimics the TN results; (ii) a nonspinning \texttt{IMRPhenomD} model including the merger and ringdown (MR); (iii) a nonspinning \texttt{IMRPhenomHM} model including both MR and higher-order modes (HM); (iv) an \texttt{IMRPhenomHM} model with extremal spins aligned with the orbital angular momentum ($\chi_1 = \chi_2 = 1$), henceforth \texttt{IMRPhenomHM+}; and (v) an \texttt{IMRPhenomHM} model with extremal spins anti-aligned with respect to the orbital angular momentum ($\chi_1 = \chi_2 = -1$), henceforth \texttt{IMRPhenomHM-}. 

Figure~\ref{fig:waveformComparison} shows how well the observation of a ``typical'' MBHB by LISA could constrain the mass of a PM lens in the redshifted-lens-mass range $\lensMass \in [10^3, 10^9]\, \Msun$.
For concreteness, we focus on a single value of the impact parameter $y=0.1$. We consider the five waveform models listed above, and we do not normalize the results by the SNR. 
The general trend with mass is similar to the findings of the inspiral-only TN analysis (see the left panel of Fig.~7 in~\cite{Takahashi:2003ix}), but our calculations allow us to quantify the effect of the merger/ringdown, higher harmonics, and spins. 
By comparing nonspinning binaries with the signal truncated at the ISCO with those including merger and ringdown (MR), we see that the inclusion of merger and ringdown leads to improvements by about one order of magnitude in the measurement of the lens mass. 
Models with higher harmonics (\texttt{IMRPhenomHM}, in green) lead to further improvements in measurement accuracy relative to models without higher harmonics (\texttt{IMRPhenomD}, in red), as expected. 
The dependence of the waveform on the angles is more pronounced when we include higher harmonics. As a consequence, the measurement errors for \texttt{IMRPhenomHM} have a larger ``spread'' around the median compared to the measurement errors for \texttt{IMRPhenomD}.
It is also well known that aligned (anti-aligned) spins typically increase (reduce) the SNR because of the orbital hang-up effect~\cite{Campanelli:2006uy}, and indeed we find that measurement errors are smallest for extremal aligned spins (\texttt{IMRPhenomHM+}, in brown) and largest for extremal anti-aligned spins (\texttt{IMRPhenomHM-}, in blue). For lens masses $\lensMass \gtrsim 10^7\, \Msun$, the geometric-optics limit is a good approximation, and $\Delta \lensMass/\lensMass$ depends solely on the SNR and the impact parameter $y$~\cite{Takahashi:2003ix}. Since in this calculation we have fixed $y$, the uncertainties in the large-$\lensMass$ regime shown in Fig.~\ref{fig:waveformComparison} are inversely proportional to the SNR of the signal.

Given the intrinsic parameters of an MBHB, we are interested in estimating the ``critical lens mass'' $\lensMass^{\rm crit}$, defined as the lowest lens mass for which we can extract information on either $\lensMass$ or $y$. We (somewhat arbitrarily) define this threshold as the lens mass corresponding to a 100\% relative uncertainty on the respective parameter.
For the MBHB considered in Fig.~\ref{fig:waveformComparison}, we find
$\critLensMass = 1.08\substack{+0.64 \\ -0.40} \times 10^5\, \Msun$ for \texttt{IMRPhenomD} truncated at the ISCO,
$\critLensMass = 4.73\substack{+2.45 \\ -1.75} \times 10^3\, \Msun$ for \texttt{IMRPhenomD},
$\critLensMass = 3.19\substack{+1.39 \\ -1.17} \times 10^3\, \Msun$ for \texttt{IMRPhenomHM},
$\critLensMass = 1.49\substack{+0.62 \\ -0.49} \times 10^3\, \Msun$ for \texttt{IMRPhenomHM+}, and
$\critLensMass = 6.63\substack{+3.46 \\ -2.24} \times 10^3\, \Msun$ for \texttt{IMRPhenomHM-}.
The quoted values correspond to the median and $68\%$ confidence interval of each critical lens mass.
We find that the critical lens mass decreases when we include the merger and higher harmonics in the waveform model, as well as for MBHBs with large aligned spins, in agreement with the trends described earlier.

We can estimate in a similar way the critical impact parameter $y^{\rm crit}$ below which we can extract information on at least one of the lens parameters.
We consider the same MBHB, but we now assume a PM lens with fixed redshifted mass $\lensMass = 10^7\, \Msun$, and we vary the impact parameter in the range $y \in [0.01, 200]$. The relative uncertainty in $y$ follows the same qualitative trends as the uncertainties in $\lensMass$ as we vary the waveform model, and the critical impact parameters are 
$y^{\rm crit} = 52.6\substack{+17.4 \\ -10.1}$ for \texttt{IMRPhenomD} truncated at the ISCO,
$y^{\rm crit} = 92.4\substack{+24.5 \\ -16.6}$ for \texttt{IMRPhenomD},
$y^{\rm crit} = 92.0\substack{+36.5 \\ -17.3}$ for \texttt{IMRPhenomHM},
$y^{\rm crit} = 114\substack{+39 \\ -25}$ for \texttt{IMRPhenomHM+}, and
$y^{\rm crit} = 86.8\substack{+21.1 \\ -19.4}$ for \texttt{IMRPhenomHM-}.

In summary: the inspiral-only waveforms used in TN lead, in general, to an overestimate of measurement uncertainties in the lens parameters relative to waveforms including also the merger and ringdown. For this reason, their results should be regarded as conservative. The merger, ringdown, and higher-order modes can significantly improve our ability to measure the lens mass, and the ``critical'' measurable lens mass $\critLensMass$ varies by a factor of $\sim 2$ or $3$ for MBHBs with large (anti)aligned spins.

\begin{figure*}[t]
    \centering
    \includegraphics[width=\textwidth]{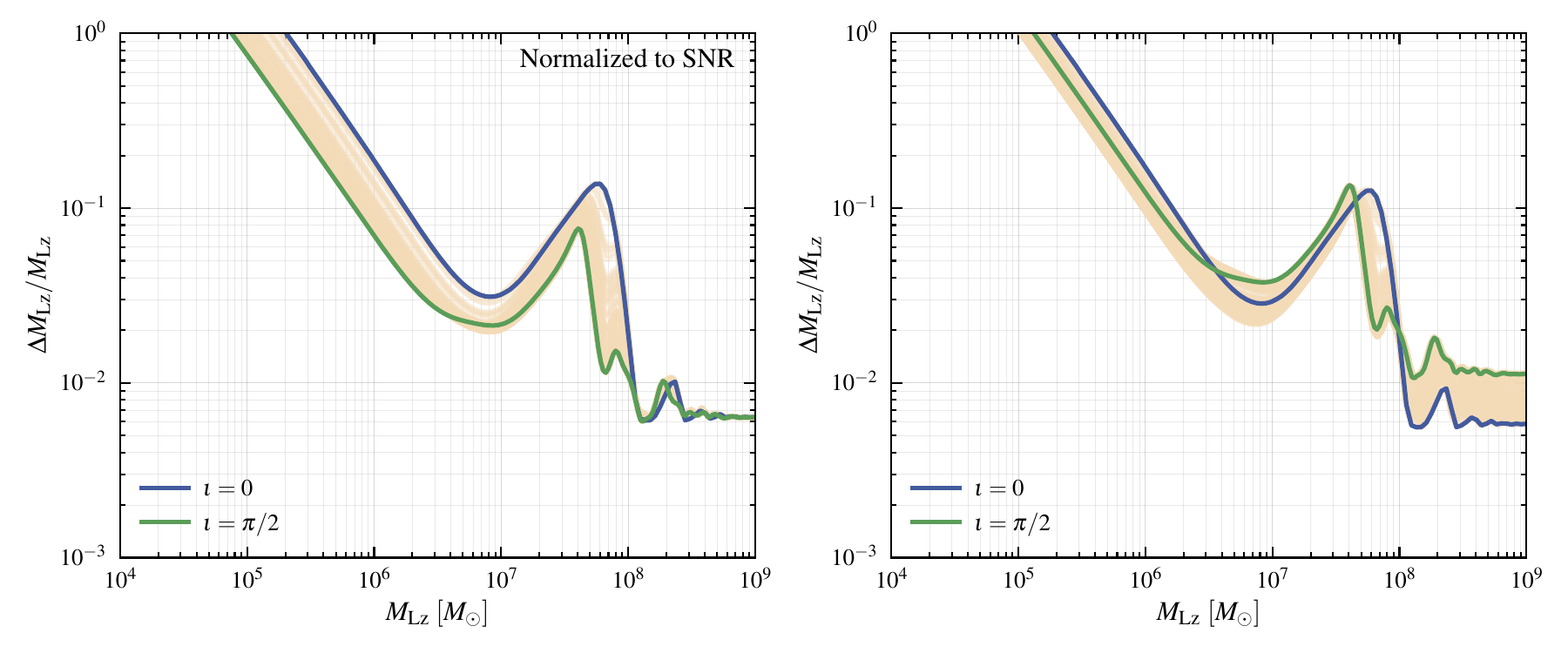}
    \caption{Effect of the MBHB inclination angle $\iota$ on the estimation accuracy of the redshifted lens mass $\lensMass$. 
    The results are for an MBHB with $M_{\rm Tz} = 10^7\, \Msun$, $q=1.2$, $z_{\rm S} = 5$, $\chi_1 = \chi_2 = 0$, and $\phi_c = \lambda = \beta = \psi = \pi/3$, and a PM lens with impact parameter $y=0.1$. We show $\orderOf(100)$ random realizations of $\iota$ (in beige) but also plot results for two selected values of $\iota = 0$ and $\pi/2$ (blue and green, respectively). 
    In the left panel, all uncertainties have been rescaled to a fixed $\textrm{SNR} = 1000$. In the right panel, this normalization was not applied.
    }%
    \label{fig:inclination}%
\end{figure*}

\subsection{Effect of the source parameters on the measurement of the lens parameters} \label{sec:EffectsOfTheSourceParameters}

We now focus on the effect of the source parameters on the measurement of $\lensMass$ and $y$. We consider \texttt{IMRPhenomHM} MBHB waveforms with 
$M_{\rm Tz} = 10^7\, \Msun$, $q=1.2$, $z_{\rm S} = 5$, and five selected values of the aligned binary component spins: $\chi_1=\chi_2 \in \{-1, -0.5, 0, 0.5, 1\}$.
To begin with, we focus on PM lenses with $y=0.1$ and $\lensMass \in [10^3, 10^9]\, \Msun$. 

We examine the effect of various parameters, namely: the inclination angle $\iota$; the magnitude of the spins; and, finally, the sky location angles (right ascension $\lambda$ and declination $\beta$), mass ratio $q$, coalescence phase $\phi_c$, and polarization angle $\psi$. 
We consider $\mathcal{O}(100)$ random values for each set of parameters that we vary and fix all angles that are not being varied to an ``intermediate'' value of $\pi/3$.

When we explore the effect of each source parameter, we either normalize the resulting errors in the lens parameters to a reference $\textrm{SNR} = 1000$, or we consider MBHBs at fixed redshift. 
This allows us to understand whether the lens parameter estimation accuracy is dominated by the SNR of the source or by more subtle features related to the specific parameter we vary. For example, higher harmonics (when detectable) can reduce correlations between parameters, and the relative importance of higher harmonics is strongly affected by the inclination of the binary.
We will now describe our findings for each parameter.

\begin{figure*}[t]
    \centering
    \includegraphics[width=\textwidth]{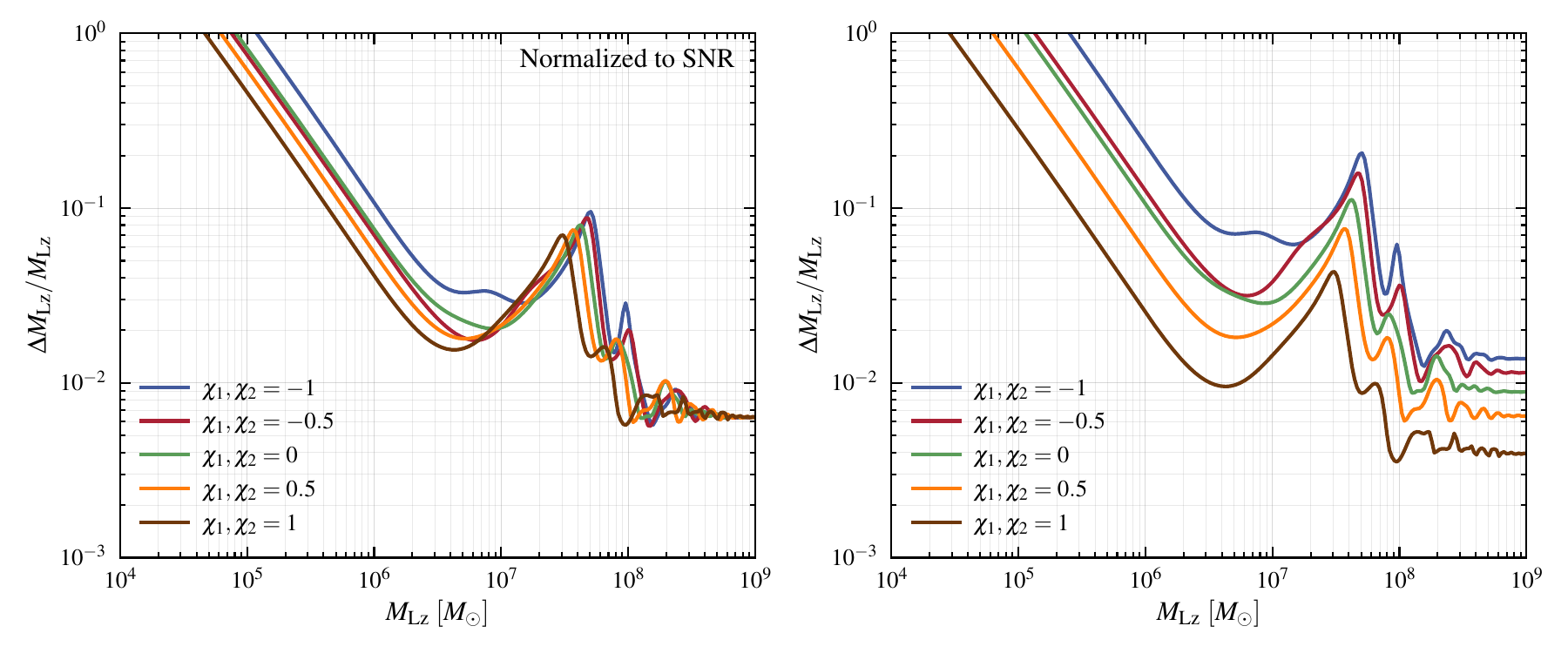}
    \caption{Effect of the MBHB spins $\chi_1$ and $\chi_2$ on $\Delta \lensMass/\lensMass$. 
    The results are for the same setting as in Fig.~\ref{fig:inclination}, but now we fix $\iota = \pi/3$, and we vary the spins in the range $\chi_1=\chi_2 = \{-1, -0.5, 0, 0.5, 1\}$ (blue, red, green, orange and brown, respectively).
    }%
    \label{fig:spin}%
\end{figure*}

\begin{figure*}[t]
    \centering
    \includegraphics[width=\textwidth]{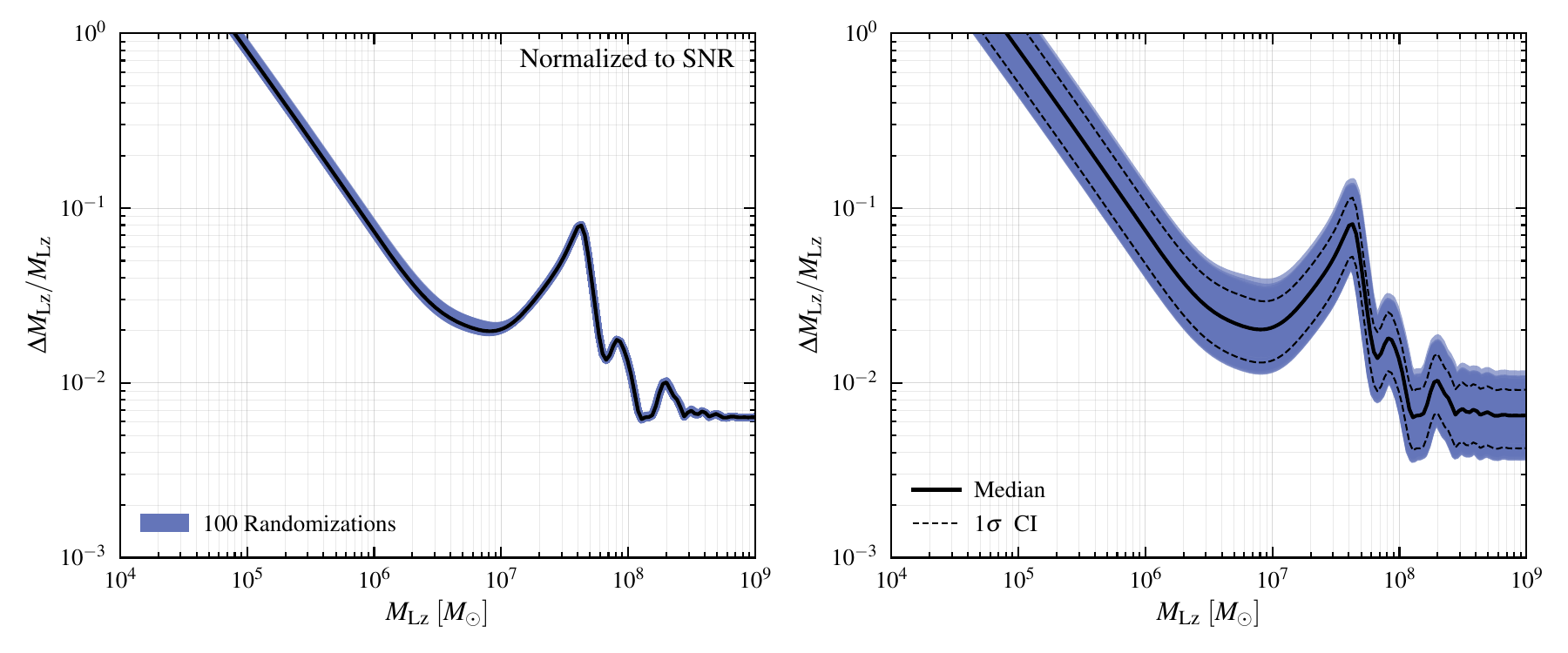}
    \caption{Effect of the MBHB sky location (right ascension and declination) on $\Delta \lensMass/\lensMass$. 
    The results are for the same setting as in Fig.~\ref{fig:inclination}, but now $\iota = \pi/3$. 
    Right panel: the median error is shown in black, the $1\sigma$ confidence interval is shown by dashed black lines, and the individual realizations are in blue.
    Left panel: we only show the median since errors are dominated by the SNR and the dispersion is minimal.
    }%
    \label{fig:skyLocation}%
\end{figure*}

\subsubsection{Inclination angle}

In Fig.~\ref{fig:inclination}, we plot $\Delta \lensMass/\lensMass$ as a function of $\lensMass$ for a sample of $\orderOf(100)$ MBHBs obtained by drawing $\iota$ uniformly in $\arccos(\iota) \in [-1, 1]$. 
In the left panel, all errors are normalized to $\textrm{SNR} = 1000$; in the right panel, the binary is located at a fixed redshift $z_{\rm S} = 5$.

Consider, first, the left panel. Face-on ($\iota = 0$) and face-off ($\iota = \pi$) binaries yield the same lens mass uncertainties, as we would expect based on symmetry, and therefore we only show errors for $\iota=0$.
Face-on and face-off binaries yield the largest errors in $\lensMass$ in the small-$\lensMass$, wave-optics regime. This is because the amplitude of higher-order modes, which are important to remove degeneracies between parameters, are suppressed for these values of $\iota$. 
Indeed, the errors are smallest for edge-on binaries ($\iota = \pi/2$), when higher-order modes matter the most.
In the large-$\lensMass$, geometric-optics regime, $\Delta \lensMass/\lensMass$ depends only on the impact parameter $y$ and on the SNR of the binary~\cite{Takahashi:2003ix}. 
Since all of our binaries have the same $\textrm{SNR}=1000$ and we fix $y=0.1$, $\Delta \lensMass/\lensMass$ tends to a constant for large $\lensMass$, as expected.

Similar trends can be observed in the right panel.
In the large-$\lensMass$ (geometric-optics) regime, the errors depend solely on the SNR of the binary because $y=0.1$ is fixed: edge-on binaries ($\iota = \pi/2$, which have the smallest SNR) yield the highest errors, while face-on and face-off binaries ($\iota = 0$ and $\iota = \pi$, which have the highest SNR) yield the smallest errors.
The situation is partially reversed in the wave-optics regime because higher-order modes remove degeneracies, partially compensating for the smaller SNR of the edge-on binaries. 

The critical redshifted lens mass above which lensing effects become detectable is $\critLensMass = 9.91\substack{+2.81 \\ -0.56} \times 10^4\, \Msun$ for fixed SNR, and $\critLensMass = 1.23\substack{+0.26 \\ -0.08} \times 10^5\, \Msun$ for MBHBs at fixed distance. Therefore, variations in the inclination angle $\iota$ lead to a relative uncertainty of $\approx 30\%$ ($\approx 20\%$) within the $1\sigma$ credible interval for binaries at fixed SNR (fixed distance, respectively).

We explored different values of the impact parameter and SIS lenses, finding qualitatively similar conclusions.
We also studied MBHBs with different total masses and mass ratios. In general, higher harmonics are more important for unequal-mass binaries, and this results in larger variances in $\critLensMass$ as we vary $\iota$.

\subsubsection{Spins}

In Fig.~\ref{fig:spin}, we show how spins affect the estimate of the lens mass. We consider five different spin combinations: $\chi_1=\chi_2 \in \{-1, -0.5, 0, 0.5, 1\}$. 

When we normalize to the SNR (left panel), large aligned (anti-aligned) spins produce lower (larger) errors in the wave-optics regime. 
All errors converge to the same value in the geometric-optics regime for the reasons explained above.
The same trend is visible and more pronounced for binaries at fixed redshift (right panel): large aligned (anti-aligned) spins produce lower (larger) errors in both the wave-optics and geometric-optics regimes.

Most of these trends are explained by the fact that aligned (anti-aligned) spins increase (reduce) the SNR because of the orbital hang-up effect~\cite{Campanelli:2006uy}.
Aligned spins affect the measurement of the lens mass even at constant SNR because the orbital hang-up effect causes the binary to spend more cycles in band and therefore reduces parameter estimation errors.

By sampling $\orderOf(100)$ MBHBs with $\chi_1, \chi_2$ uniformly distributed in the range $[-1, 1]$, while keeping all other parameters fixed,
we find a critical redshifted lens mass of
$\critLensMass = 7.63\substack{+2.04 \\ -1.81} \times 10^4\, \Msun$ (relative uncertainty of $\approx 25\%$) for fixed SNR, and
$\critLensMass = 1.15\substack{+0.67 \\ -0.63} \times 10^5\, \Msun$ (relative uncertainty of $\approx 50\%$) for fixed distance.

Qualitatively, we find similar results when we vary the impact parameter, consider the SIS lens model, or change the MBHB masses.

\subsubsection{Sky location, mass ratio, coalescence phase, and polarization angle}

In Fig.~\ref{fig:skyLocation}, we consider $\orderOf(100)$ MBHBs with sky location (right ascension and declination) uniformly distributed on the celestial sphere. 
The minimal dispersion of the uncertainties seen in the left panel shows that while sky location affects $\Delta \lensMass/\lensMass$, the effect is predominantly due to the different SNR of binaries located at different positions in the sky. 
The critical redshifted lens mass $\critLensMass = 8.35\substack{+3.93 \\ -3.03}  \times 10^4\, \Msun$ (with a relative uncertainty of $\approx 45\%$) when the redshift is fixed; the median is the same (but with a minimal relative uncertainty $\lesssim 1\%$) when we fix the SNR.

To understand the effect of varying mass ratio $q$, we varied $q$ uniformly in the range $[1, 10]$ $\orderOf(100)$ times.
We found $\critLensMass = 1.87\substack{+0.97 \\ -0.73}  \times 10^5\, \Msun$ (with a relative uncertainty of $\approx 50\%$) at fixed distance, and $\critLensMass = 7.34\substack{1.01 \\ -0.46}  \times 10^4\, \Msun$ (with a relative uncertainty of approximately $\approx 15\%$) at fixed SNR. As expected, varying the mass ratio affects how pronounced the higher-order modes are, which can lower measurement uncertainties. However, changing the mass ratio also affects the SNR of the signal. Therefore, both the effects (degeneracy removal by higher-order modes versus reduced SNR) affect the result when the SNR rescaling is not applied. This is similar to the case of varying the inclination angle.

We also varied the coalescence phase in the range $\phi_c \in [0, 2\pi]$. We found $\critLensMass = 1.08\substack{+0.15 \\ -0.07}  \times 10^5\, \Msun$ (with a relative uncertainty of $\approx 10\%$) at fixed distance, and an even smaller uncertainty ($\lesssim 6\%$) at fixed SNR. 
By varying the polarization angle uniformly in the range $\psi \in [0, 2\pi]$ we find $\critLensMass = 1.23\substack{+0.08 \\ -0.08}  \times 10^5\, \Msun$ ($\approx 6\%$ uncertainty), with
an even smaller uncertainty ($\lesssim 5\%$) at fixed SNR. 

Once again, the results are qualitatively similar when we vary the impact parameter, consider the SIS lens model, or change the MBHB masses.

\begin{figure*}[t]
    \centering
    \includegraphics[width=.97\textwidth]{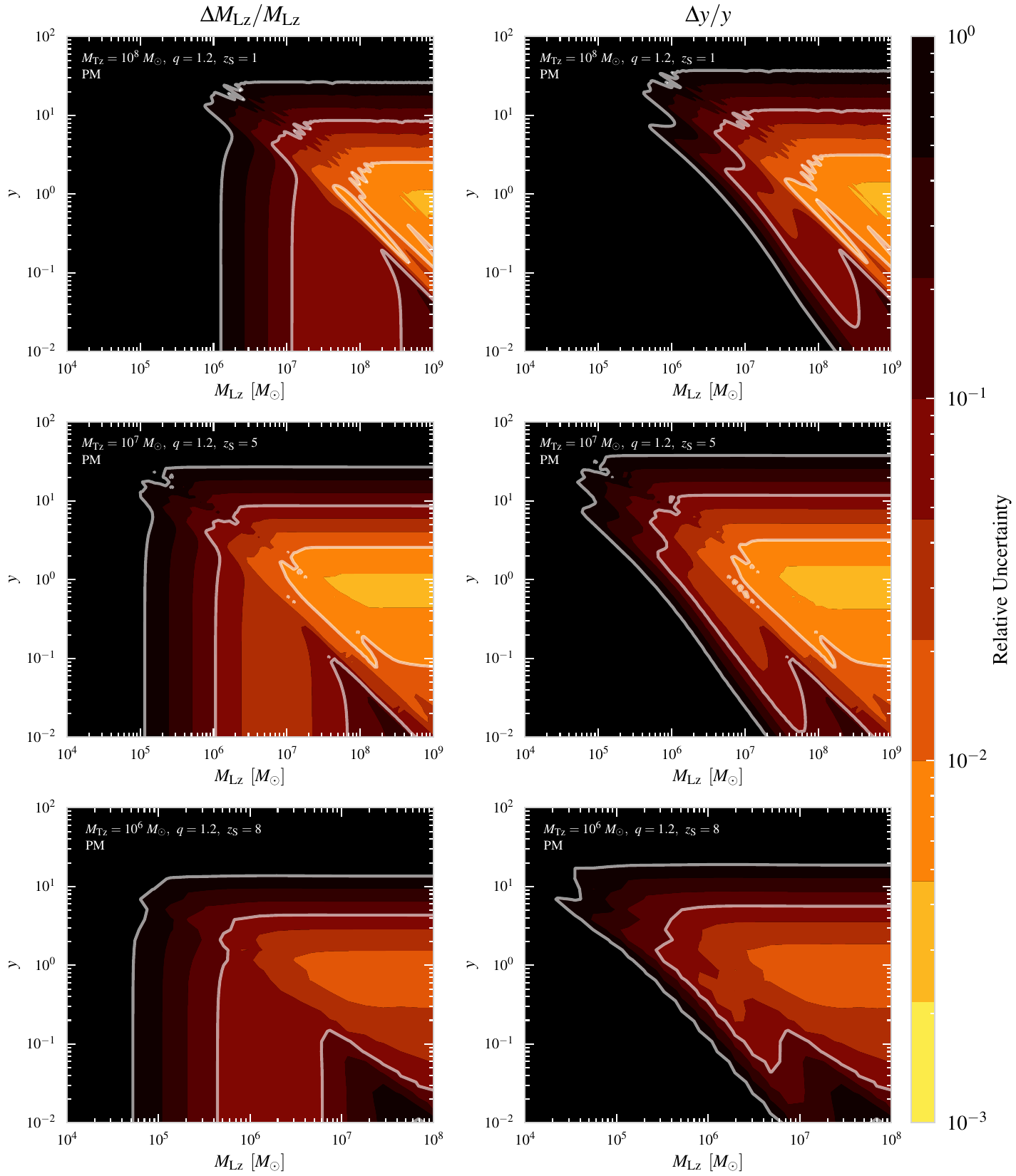}%
    \caption{Relative errors in the redshifted lens mass $\Delta \lensMass/\lensMass$ (left panels) and impact parameter $\Delta y/y$ (right panels) in the $(M_{\rm Lz},\,y)$ plane for a PM lens. 
    The rows refer to three different MBHBs, with parameters listed in the legend. White contour lines correspond to 100\%, 10\%, and 1\% relative errors. In the black regions, the relative errors are larger than 100\%, and the corresponding parameter is unmeasurable.
    The MBHBs' unlensed $\textrm{SNR}$ is $697.0,\, 715.8,\,$ and $181.8$ from top to bottom.
    }%
    \label{fig:PML_MP}%
\end{figure*}

\begin{figure*}[t]
    \centering
    \includegraphics[width=.97\textwidth]{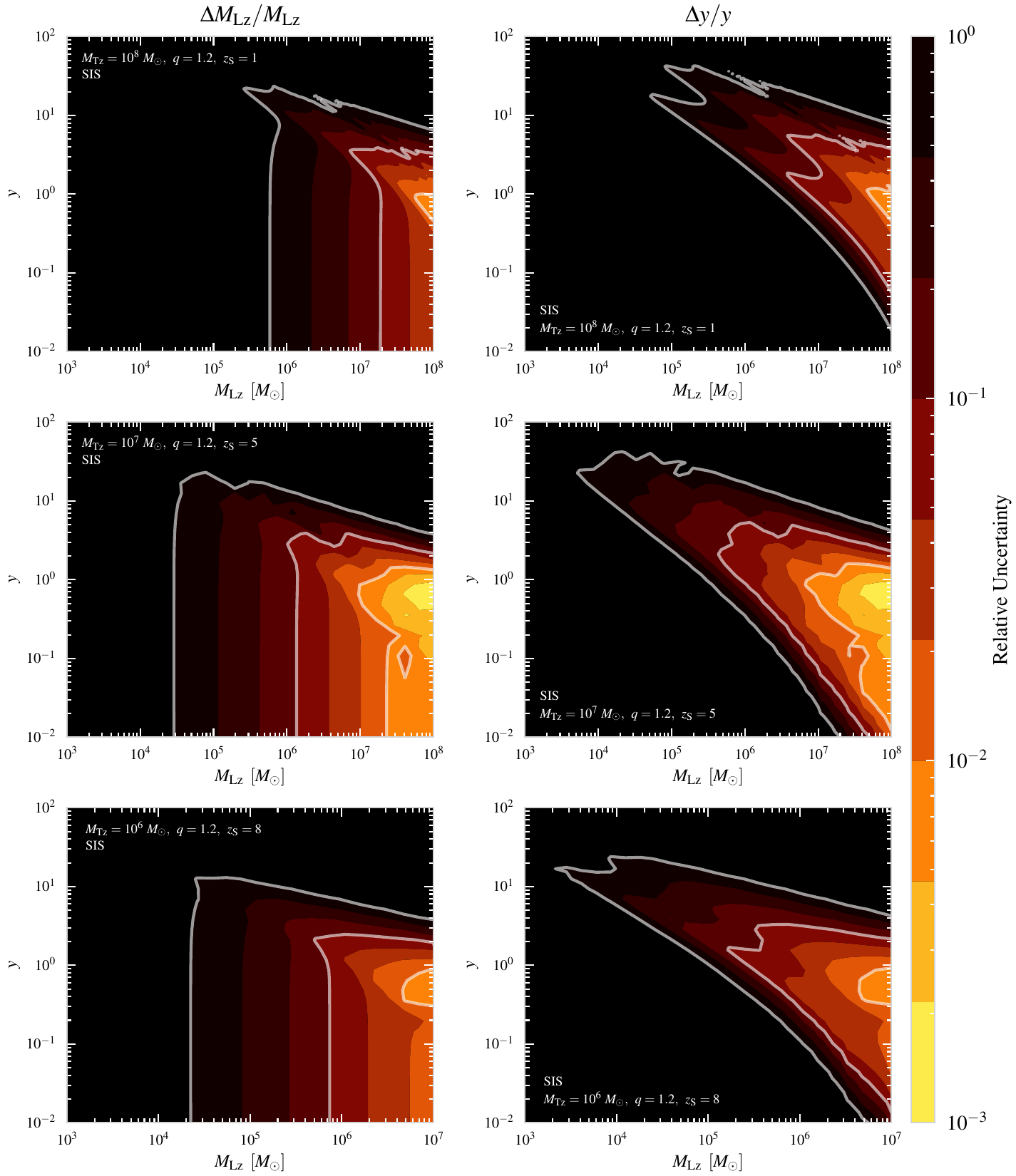}%
    \caption{Same as Fig.~\ref{fig:PML_MP}, but for an SIS lens.}%
    \label{fig:SIS_MP}%
\end{figure*}

\subsection{Point-mass lens} \label{sec:Results_PML}

So far, we have investigated how the measurement of lens parameters is affected by waveform modeling and source parameters.
We will now consider three representative MBHBs and compute lens parameter estimation accuracy for a wide range of lens masses and impact parameters.
In this section, we focus on PM lenses with $y \in [0.01, 200]$ and $\lensMass \in [10^3, 10^9]\, \Msun$.

In Fig.~\ref{fig:PML_MP}, we show contour plots of $\Delta \lensMass/\lensMass$ (left panels) and $\Delta y/y$ (right panels) in the $(\lensMass,\,y)$ plane.
Different rows refer to three different nonspinning MBHBs with
detector-frame mass $M_{\rm Tz} = 10^8\, \Msun$, $q=1.2$, $z_{\rm S}=1$ (top); 
$M_{\rm Tz} = 10^7\, \Msun$, $q=1.2$, $z_{\rm S}=5$ (middle); and
$M_{\rm Tz} = 10^6\, \Msun$, $q=1.2$, $z_{\rm S}=8$ (bottom). These masses and redshifts have been chosen as representative of typical MBHB systems observable by LISA (see, e.g.,~\cite{Sesana:2010wy,Klein:2015hvg,Toubiana:2021iuw}).
To reduce computational time, the angles $\iota$, $\phi_c$, $\lambda$, $\beta$ and $\psi$ were all set to $\pi/3$. The range of variability of the results around these ``intermediate'' values was discussed in Sec.~\ref{sec:EffectsOfTheSourceParameters} above.

The SNRs of the unlensed signals from these MBHBs are $ 697,\, 716,$ and $182$ for the top, middle, and bottom panels in Fig.~\ref{fig:PML_MP}, respectively.
When the signals are lensed, the SNRs increase up to $\sim 5970$, $\sim 9180$, and $\sim 2050$ for the top, middle, and bottom panels, respectively.

Three white contour lines in each panel highlight the 100\%, 10\%, and 1\% relative uncertainty boundary regions. In the black regions (outside the outermost white contour), the relative uncertainty is greater than 100\%, and therefore at least one of the lensing parameters is unmeasurable. In fact, in some regions of the parameter space, we can measure only one of the lens parameters.

Consider, for example, an MBHB with $M_{\rm Tz} = 10^8\, \Msun$ at $z_{\rm S}=1$ (top row) with lens parameters $\lensMass \approx 10^7\, \Msun$ and $y \approx 0.1$: in this case the lens mass can be measured with $\approx 10\%$ relative uncertainty, but $y$ is unmeasurable.
For this same binary, $\lensMass$ is measurable when $y\lesssim 30$ and $\lensMass \gtrsim 10^6\, \Msun$ (top left panel), while $y$ is measurable when $y\lesssim 40$ (top right panel).
These ``detectability boundaries'' are slightly different for lighter binaries. We show representative examples in Table~\ref{tab:PML_results}.
Based on this extensive analysis, we conclude that the critical values of $\critImpactParam$ and $\critLensMass$ are more optimistic than the TN predictions for PM lenses because the merger/ringdown and higher-order modes sensibly reduce the errors in the lens parameters.

\begin{table}[t]
\renewcommand{\arraystretch}{1.5}
\caption{\label{tab:PML_results}%
Four binaries of various redshifted total masses (first column), mass ratio (second column), and redshift (third column), we list: the lowest redshifted lens mass $\critLensMass$ and highest impact parameter $\critImpactParam$ for which $\lensMass$ is measurable (fourth and fifth columns); and the largest impact parameter $\critImpactParam$ for which $y$ is measurable (sixth column). 
The first three rows refer to the binaries considered in Fig.~\ref{fig:PML_MP}; the fourth row refers to the binary shown in Fig.~\ref{fig:waveformComparison} (where we fix the angles $\iota$, $\phi_c$, $\lambda$, $\beta$, and $\psi$ to $\pi/3$).
All results in this table are for PM lenses.
}
\begin{ruledtabular}
\begin{tabular}{lccllll}
                &     &   & \multicolumn{2}{c}{$\Delta \lensMass/\lensMass$} & \multicolumn{2}{c}{$\Delta y/y$} \\ \cline{4-7} 
\multicolumn{1}{c}{$M_{\rm Tz}\ [\Msun]$} & \multicolumn{1}{c}{$q$} & \multicolumn{1}{c}{$z_{\rm S}$} & \multicolumn{1}{c}{$\critLensMass\ [\Msun]$} & $\critImpactParam$ & \multicolumn{2}{c}{$\critImpactParam$} \\
\colrule
$10^8$          & 1.2 & 1 & $\gtrsim10^6$           & $\lesssim30$  & \multicolumn{2}{c}{$\lesssim40$}      \\
$10^7$          & 1.2 & 5 & $\gtrsim10^5$          & $\lesssim30$  & \multicolumn{2}{c}{$\lesssim40$}      \\
$10^6$          & 1.2 & 8 & $\gtrsim 5 \times10^4$   & $\lesssim15$  & \multicolumn{2}{c}{$\lesssim20$}      \\
$2 \times 10^6$ & 1   & 1 & $\gtrsim 4 \times 10^3$              & $\lesssim 60$             & \multicolumn{2}{c}{$\lesssim 80$}     
\end{tabular}
\end{ruledtabular}
\label{tab:criticalpars}
\end{table} 

\subsection{Singular isothermal sphere lens} \label{sec:Results_SIS}

We now consider the same MBHBs as in Sec.~\ref{sec:Results_PML}, but we extend the analysis to SIS lenses with $y\in[0.01, 200]$ and $\lensMass \in [10^1, 10^8]\, \Msun$. 
In Fig.~\ref{fig:SIS_MP}, we show contour plots of $\Delta \lensMass/\lensMass$ (left panels) and $\Delta y/y$ (right panels) in the $(\lensMass, y)$ plane. Once again, we set all angles ($\iota$, $\phi_c$, $\lambda$, $\beta$, $\psi$) equal to $\pi/3$.

The SNRs of the unlensed signals were listed in Sec.~\ref{sec:Results_PML}.
When the signals are lensed, the SNRs increase up to $\sim 2620$, $\sim 7660$, and $\sim 1180$ for the top, middle, and bottom MBHBs in Fig.~\ref{fig:SIS_MP}.

For the binary with $M_{\rm Tz} = 10^8\, \Msun$ (top row), $\lensMass$ is measurable when $y \lesssim 20$ and $\lensMass \gtrsim 6\times 10^5\, \Msun$, while $y$ is measurable when $y \lesssim 45$.
In some regions of the parameter space, we can measure only one of the lens parameters.
A qualitative difference with respect to PM lenses is that $\critImpactParam$ is no longer (approximately) constant, but it depends on $\lensMass$ (as expected from, e.g., Fig.~10 of TN).
We summarize the results for each binary in Table~\ref{tab:SIS_results}.

For the binary with $M_{\rm Tz} = 10^7\, \Msun$ (middle row), we find that the highest $\critImpactParam$ corresponds to $\lensMass \approx 10^4\, \Msun$. Ref.~\cite{Gao:2021sxw} found the maximum value of the inner product defining the Lindblom criterion, $\langle \delta h | \delta h \rangle \approx 6$, occurs for a comparable value of $\lensMass$, and that $\langle \delta h | \delta h \rangle$ decreases -- while still satisfying the condition $\langle \delta h | \delta h \rangle > 1$ -- for lower values of $\lensMass$.
Note, however, that, according to our analysis, none of the lens parameters is measurable for $\lensMass \lesssim 3 \times 10^3\, \Msun$. This implies that the Lindblom criterion is necessary but not sufficient to conclude whether lensing is observable.

Our findings are significantly more optimistic than those in TN: by including the merger, ringdown, and higher harmonics, we can measure lens parameters for higher values of $y$ and lower values of $\lensMass$ than previously thought.
Even if we consider a more stringent measurability criterion (setting the threshold at, say, 10\% relative uncertainty), we still find that we can extract information about the lens parameters for higher values of $y$ and lower values of $\lensMass$ than estimated by TN. 
As in the case of PM lenses, the values of $\critLensMass$ and $\critImpactParam$ have a strong dependence on the parameters of the source and the lens. The simple estimates of lensing probability by TN assumed $\critImpactParam$ to be constant, but a more careful estimate should consider the dependence of $\critLensMass$ or $\critImpactParam$ on the source and lens parameters.

\begin{table}[t]
\renewcommand{\arraystretch}{1.5}
\caption{\label{tab:SIS_results}%
Same as Table~\ref{tab:PML_results}, but for an SIS lens.
}
\begin{ruledtabular}
\begin{tabular}{lccllll}
                &     &   & \multicolumn{2}{c}{$\Delta \lensMass/\lensMass$} & \multicolumn{2}{c}{$\Delta y/y$} \\ \cline{4-7} 
\multicolumn{1}{c}{$M_{\rm Tz}\ [\Msun]$} & \multicolumn{1}{c}{$q$} & \multicolumn{1}{c}{$z_{\rm S}$} & \multicolumn{1}{c}{$\critLensMass\ [\Msun]$} & $\critImpactParam$ & \multicolumn{2}{c}{$\critImpactParam$} \\
\colrule
$10^8$          & 1.2 & 1 & $\gtrsim6\times 10^5$ & $\lesssim20$    & \multicolumn{2}{c}{$\lesssim45$}      \\
$10^7$          & 1.2 & 5 & $\gtrsim3\times 10^4$ & $\lesssim25$    & \multicolumn{2}{c}{$\lesssim40$}      \\
$10^6$          & 1.2 & 8 & $\gtrsim 2\times 10^4$   & $\lesssim 10$    & \multicolumn{2}{c}{$\lesssim 20$}      
\end{tabular}
\end{ruledtabular}
\end{table}

\section{Conclusions and outlook} \label{sec:Conclusions}

Wave-optics effects in lensed GW signals emitted by MBHBs in the LISA band are important when the Schwarzschild radius of the lens is smaller than the wavelength of radiation
[cf.\ Eq.~(\ref{eqn:diffraction_condition})].
If detected, these frequency-dependent wave-optics effects could lead to a plethora of applications, such as precision cosmology or constraints on the population of lenses.

We have studied the observability of wave-optics effects by LISA. We computed the parameter-estimation errors using analytical solutions for both PM and SIS lenses. These analytical solutions allow us to compute the derivatives of the diffraction integral $F(w,y)$.
In the context of lensing, this is (to our knowledge) the first study using gravitational-waveform models that include the merger, ringdown, higher harmonics, and aligned spins.
We found that the inspiral-only waveforms used in previous work overestimate measurement uncertainties in the lens parameters by about an order of magnitude.
The merger, ringdown, and higher-order modes significantly improve our ability to measure $\lensMass$ and $y$.
The ``critical'' value of the redshifted lens mass for which such measurements are possible varies by a factor of $\sim 2$ or $3$ for MBHBs with large (anti)aligned spins.

We selected three representative MBHBs that could be detectable by LISA and performed an extensive parameter estimation survey for a wide range of lens masses and impact parameters.
The results for PM (SIS) lenses are shown in Fig.~\ref{fig:PML_MP} (Fig.~\ref{fig:SIS_MP}) and Table~\ref{tab:PML_results} (Table~\ref{tab:SIS_results}). We found that the critical values of the lens mass and impact parameter for which lensing is measurable depend very strongly on the source parameters. Therefore, assuming these critical parameters to be constant can lead to incorrect estimates of the lensing probability.

As claimed by Ref.~\cite{Gao:2021sxw}, the lens parameters could be measurable for SIS lenses with impact parameters satisfying $y>3$. However, (contrary to the claims of Ref.~\cite{Gao:2021sxw}) we found that none of the SIS lens parameters are measurable for $\lensMass \lesssim 3 \times 10^3\, \Msun$: this shows that the Lindblom criterion is not accurate enough to determine whether lensing is observable.

Our parameter-estimation study shows that GW lensing of MBHBs with LISA may be more easily observable than previously thought. Estimating the rate of observable lensing events requires population studies based on astrophysical models~\cite{Sesana:2010wy,Klein:2015hvg,Toubiana:2021iuw}, and it is an exciting topic for future work.

\acknowledgements
We are grateful to Neha Anil Kumar, Selim C.\ Hotinli, and Jose M.\ Ezquiaga for insightful discussions and feedback on this manuscript.
We thank Jaime Combariza for his help and guidance in high-performance computing.
We also thank Ryuichi Takahashi for bringing to our attention the analytical Taylor series solution of the SIS lensing diffraction integral presented in Ref.~\cite{Matsunaga:2006uc}.
M.\c{C}., R.C., and E.B.\ are supported by NSF Grants No.\ AST-2006538, PHY-2207502, PHY-090003 and PHY20043, and NASA Grants No. 19-ATP19-0051, 20-LPS20- 0011 and 21-ATP21-0010. M.K.\ and L.J.\ were supported by NSF Grant No.\ 1818899 and the Simons Foundation.
M.\c{C}.\ is also supported by Johns Hopkins University through the Rowland Research Fellowship.
This work was carried out at the Advanced Research Computing at Hopkins (ARCH) core facility (\url{rockfish.jhu.edu}), which is supported by the NSF Grant No. OAC-1920103.
The authors acknowledge the Texas Advanced Computing Center (TACC) at The University of Texas at Austin for providing {HPC, visualization, database, or grid} resources that have contributed to the research results reported within this paper \cite{10.1145/3311790.3396656}. URL: \url{http://www.tacc.utexas.edu}

\appendix

\begin{figure*}[t]
    \centering
    \includegraphics[width = \linewidth]{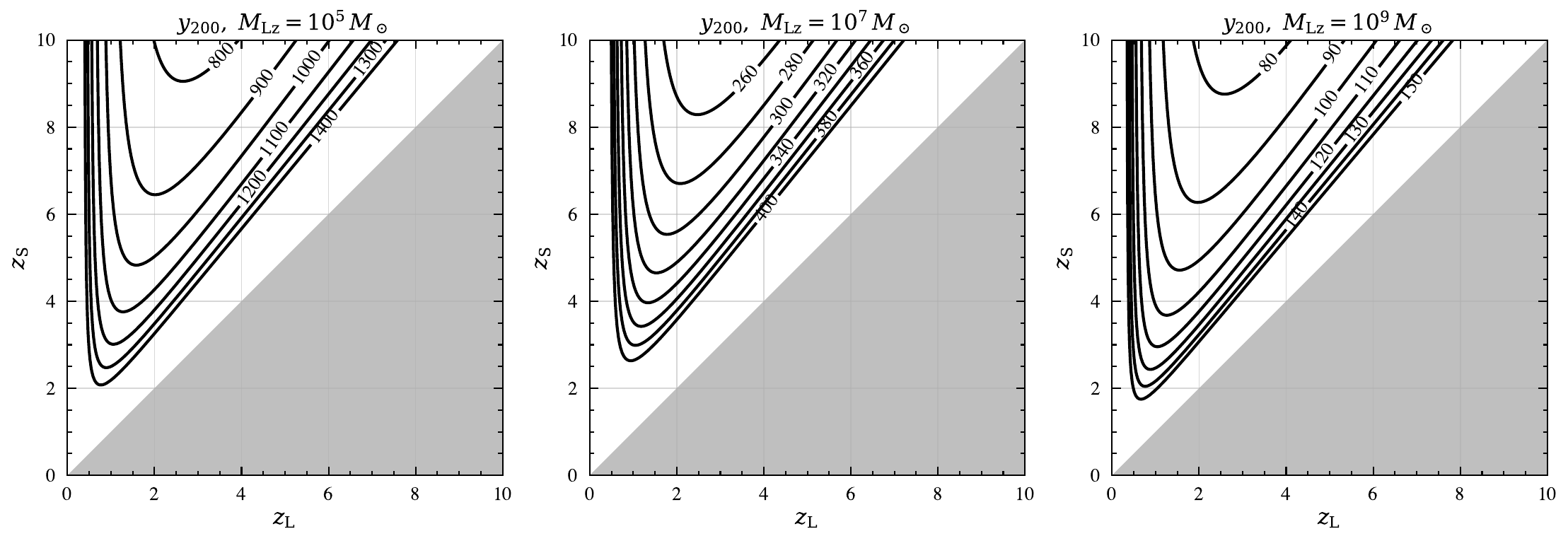}%
    \caption{The impact parameter $y_{200}$ corresponding to the $r_{200}$ boundary as a function of the lens redshift $z_\mathrm{L}$ and the source redshift $z_\mathrm{S}$ for three redshifted lens masses (left to right): $M_\mathrm{Lz} = \{10^5, 10^7, 10^9\} \, M_\odot$.}
    \label{fig:y-200}
\end{figure*}

\section{Evaluation of $F(w, y)$ for Singular Isothermal Sphere Lenses}\label{sec:SIS-evaluation}

We evaluate $F(w, y)$ for an SIS lens using a piecewise strategy: we sum the series in Eq.~\eqref{eqn:amplification-factor-spherical-series} for small $(w, y)$, and we use the geometric-optics approximation for large $(w, y)$. More in detail, we set
\begin{equation}
    F(w, y) = \begin{cases}
    F_\mathrm{wave}(w, y) & (w < 250 \land wy < 250), \\
    F_\mathrm{geom}(w, y) & (\text{otherwise}).
    \end{cases}
\end{equation}
Here, $F_\mathrm{wave}(w, y)$ denotes the right-hand side of Eq.~\eqref{eqn:amplification-factor-spherical-series}, and $F_\mathrm{geom}(w, y)$ is the geometric-optics approximation:
\begin{equation}
    F_\mathrm{geom}(w, y) = \begin{cases}
        |\mu_+|^{1/2} - i |\mu_-|^{1/2} \exp(2i wy) & (y\leq 1), \\ 
        |\mu_+|^{1/2} & (y > 1),
    \end{cases}
\end{equation}
where $\mu_\pm = \pm 1 + 1/y$ is the magnification of the images in the geometric-optics limit.

When numerically evaluating $F_\mathrm{wave}(w, y)$, we use the \texttt{nsum} function from the package for real and complex floating-point arithmetic with arbitrary precision \texttt{mpmath v1.2.1}~\cite{mpmath},
with a floating-point precision parameter \texttt{dps = 50} and the options \texttt{\{`method':`shanks',`tol':1e-15\}}. We have verified that these options can ensure convergence of the series, as well as the convergence of the series expansions required to evaluate the derivatives (see Appendix~\ref{sec:App_Derivatives_SIS}), in the $(w,y)$ region relevant for our study. When using the geometric-optics approximation $F_\mathrm{geom}(w, y)$, we have also verified that the error is always $\lesssim 10\%$.

\section{Singular Isothermal Sphere with an Outside Boundary}\label{sec:SIS-boundary}

In the main text, we noted that the total mass of the SIS profile is formally infinite and that this divergence is usually regularized by introducing an outer boundary at $r = r_\Delta$ such that $\rho_\mathrm{SIS}(r_\Delta) = \Delta \rho_\mathrm{cr}$, where $\Delta$ is a dimensionless constant. Solving for $r_\Delta$ as defined above yields $ r_\Delta = [4/(3\Delta)]^{1/2} \sigma_v / H_\mathrm{L}$, which leads to a normalized lens-plane coordinate $x_\Delta \equiv r_\Delta / \xi_0$. For the lensing configurations relevant to this paper, $x_\Delta$ is in the single-image regime. We can thus compute the corresponding impact parameter $y_\Delta = x_\Delta - 1$, with the result
\begin{equation}
    y_\Delta = \frac{1}{2\pi \sqrt{3\Delta}} \left(\frac{1}{\sigma_v H_\mathrm{L}}\right) \left(\frac{D_\mathrm{S}}{D_\mathrm{L} D_\mathrm{LS}}\right) - 1.
\end{equation}
The velocity dispersion $\sigma_v$ can then be related to the redshifted lens mass $M_\mathrm{Lz}$ using Eq.~(\ref{eq:MLsigmav}). In Fig.~\ref{fig:y-200}, we plot $y_{200}$ as a function of $z_\mathrm{L}$ and $z_\mathrm{S}$ for three selected values of $M_\mathrm{Lz}$. We conclude that, for the lensing configurations relevant to this paper, the impact parameter $y_{200}$ corresponding to the $r_{200}$ boundary is greater than $\sim 100$, and thus the truncation of the SIS profile is irrelevant in the range of potentially detectable values of $y$.

\section{Analytical Derivatives}
\label{app:analytic_derivatives}
Equations~\eqref{eq:lensed_modes} and~\eqref{eq:reduced_TDI} imply that the reduced TDI observables $\Tilde{X}^L$ of a lensed MBHB have the form
\begin{equation} \label{eqn:lensed_waveform_theta}
    \Tilde{X}^{\rm L}(\bm{\theta}^{\rm L}, \bm{\theta}^{\rm S}) = F(w,y) \Tilde{X}(\bm{\theta}_{\rm S})\,.
\end{equation} 
From the analytical expressions of $F(w,y)$ for the PM and SIS lenses, we can get analytical expressions for the derivatives of the lensed waveform appearing in the information matrix as follows. 

Using Eq.~\eqref{eqn:lensed_waveform_theta}, the partial derivative of the lensed waveform with respect to any parameter $\gamma$ reads
\begin{multline}\label{eq:derivative_general}
  \frac{\partial \Tilde{X}^{\rm L}(\bm{\theta}^{\rm L}, \bm{\theta}^{\rm S})}{\partial \gamma} = \left ( \frac{\partial F(w,y)}{\partial w}\cdot \frac{\partial w}{\partial \gamma} + \frac{\partial F(w,y)}{\partial y}\cdot \frac{\partial y}{\partial \gamma} \right )\\ \times \Tilde{X}(\bm{\theta}^{\rm S}) 
  +  F(w,y) \frac{\partial \Tilde{X}(\bm{\theta}^{\rm S})}{\partial \gamma}. 
\end{multline}
If $\gamma \in \bm{\theta}^{\rm L}$, all terms proportional to $\partial \tilde{X}(\bm{\theta}^{\rm S})/\partial \gamma$ vanish. Similarly, if $\gamma \in \bm{\theta}^{\rm S}$, all partial derivatives of $F(w,y)$ with respect to $\gamma$ vanish. We compute the numerical derivatives with respect to the source parameters using the software \texttt{lisabeta}~\cite{Marsat:2018oam}. The derivatives with respect to the lens parameters are computed below, first for PM lenses and then for SIS lenses.

In summary: the derivatives of the lensed waveforms with respect to $\lensMass$ involve Eqs.~\eqref{eqn:derivative_gamma} and~\eqref{eqn:derivative_1F1_w}
for PM lenses, and Eqs.~\eqref{eqn:derivative_F_SIS_w} for SIS lenses. The derivatives with respect to $y$ are given by Eq.~\eqref{eqn:derivative_1F1_y} for PM lenses, and~\eqref{eqn:derivative_F_SIS_y} for SIS lenses.
Once these derivatives are known, we can use Eqs.~\eqref{eqn:information_matrix_summation} and~\eqref{eqn:variance-covariance} to estimate the errors in any of the 13 source and lens parameters $\gamma$.

\subsection{Point-mass Lens}\label{sec:App_Derivatives_PML}

The diffraction integral $F(w,y)$ depends only on the lens parameters $\lensMass$ and $y$.
From Eq.~\eqref{eqn:F_point_mass}, we see that we need derivatives of the gamma function $\Gamma(z)$ with respect to $w$, and derivatives of the confluent hypergeometric function $_{1}F_{1}(a,b;z)$ with respect to both $w$ and $y$. 
The derivatives of the other terms in Eq.~\eqref{eqn:F_point_mass} are trivial.
The gamma function in the Weierstrass form~\cite{KrantzStevenG.StevenGeorge1999Hocv} can be written as
\begin{equation}
    \Gamma(z) = \left\{ z\ e^{c z} \prod_{r = 1}^{\infty}\left[\left(1+\frac{z}{r}\right) e^{-z/r}\right] \right\}^{-1}\,,
\end{equation}
where $c$ is the Euler-Mascheroni constant~\cite{KrantzStevenG.StevenGeorge1999Hocv}, and $z \in \mathbb{C}$. Differentiating, we get
\begin{equation}
    \Gamma'(z) = \Gamma(z)\ \Psi(z),
\end{equation}
where $\Psi(z)$ is the digamma function~\cite{gradshteyn2007}.

Now, set $z=1-wi/2$ to find
\begin{equation}\label{eqn:derivative_gamma}
    \frac{\partial \Gamma\left ( 1-\frac{w}{2}i \right ) }{\partial w} = -\frac{1}{2}i\ \Gamma \left(1-\frac{w}{2}i \right)\ \Psi \left(1-\frac{w}{2}i \right)\,.
\end{equation}

The confluent hypergeometric function can be written as
\begin{equation}\label{eqn:hypergeometric_def}
    _{1}F_{1}(a,b;z) = 1 + \frac{a}{b}\frac{z}{1!}+ \frac{a(a+1)}{b(b+1)}\frac{z^2}{2!}+ ...\,.
\end{equation}
Since $b=1$, the only required derivatives are those with respect to $a$ and $z$. The partial derivative of Eq.~\eqref{eqn:hypergeometric_def} with respect to $z$ is given by
\begin{equation}\label{eqn:1F1_z}
    \frac{\partial\ _{1}F_{1}(a,b;z)}{\partial z} = \frac{a}{b}\ _{1}F_{1}(a+1,b+1;z)\,.
\end{equation}
The partial derivative of Eq.~\eqref{eqn:hypergeometric_def} with respect to $a$ is given by
\begin{multline}\label{eqn:1F1_a}
    \frac{\partial\ _{1}F_{1}(a,b;z)}{\partial a} = \sum_{k=0}^{\infty}\frac{(a)_{k}\Psi(a+k)z^{k}}{k! (b)_{k}}\\ - \Psi(a) _{1}F_{1}(a,b;z)\,,
\end{multline}
where $(a)_{k} = \Gamma(a+k)/\Gamma(a)$ is the Pochhammer symbol, and $\Psi(z)$ is the digamma function as before.

Using Eqs.~\eqref{eqn:1F1_z} and~\eqref{eqn:1F1_a}, we can find the partial derivative of $_{1}F_{1}(a,b;z)$ with respect to $w$:
\begin{widetext}
\begin{align}
\frac{\partial\ _{1}F_{1}(a,b;z)}{\partial w} &= \frac{\partial\  _{1}F_{1}(a,b;z)}{\partial a} \cdot \frac{\partial a}{\partial w} + \frac{\partial\ _{1}F_{1}(a,b;z)}{\partial z} \cdot \frac{\partial z}{\partial w} \nonumber\\
    &= \frac{i}{2}\cdot \frac{\partial\ _{1}F_{1}(a,b;z)}{\partial a} + \frac{y^2}{2}i \cdot \frac{\partial\ _{1}F_{1}(a,b;z)}{\partial z} \nonumber\\
    &= \frac{i}{2} \sum_{k=0}^{\infty}\frac{(i\frac{w}{2})_{k}\Psi(i\frac{w}{2}+k)z^{k}}{k! (1)_{k}} - \Psi\left(i\frac{w}{2}\right)\ _{1}F_{1}\left(i\frac{w}{2},1;i\frac{wy^2}{2}\right) - \frac{wy^2}{4}\ _{1}F_{1}\left(i\frac{w}{2}+1,2;i\frac{wy^2}{2}\right)\,. \label{eqn:derivative_1F1_w}
\end{align}
\end{widetext}
Then, $\partial F(w,y)/\partial w$ can be obtained by using Eqs.~\eqref{eqn:derivative_gamma} and~\eqref{eqn:derivative_1F1_w} and applying the product rule. 
Setting $w = 8\pi \lensMass f$ and using the chain rule gives $\partial F(w,y)/\partial \lensMass$.

Similarly, the partial derivative of $_{1}F_{1}(a,b;z)$ with respect to $y$ is given by
\begin{equation}\label{eqn:derivative_1F1_y}
\begin{split}
    \frac{\partial\ _{1}F_{1}(a,b;z)}{\partial y} &= w y i \cdot \frac{\partial\ _{1}F_{1}(a,b;z)}{\partial z}\\
    &= -\frac{w^2 y}{2}\ _{1}F_{1}\left(\frac{w}{2}i+1,2;\frac{w y^2}{2}i\right)\,.
\end{split}
\end{equation}
Then, $\partial F(w,y)/\partial y$ can be obtained using Eq.~\eqref{eqn:derivative_1F1_y} and applying the product rule.

\subsection{Singular isothermal sphere lens}\label{sec:App_Derivatives_SIS}

The diffraction integral $F(w,y)$ for the singular isothermal sphere lens can be analytically obtained by the perturbative expansion described in Sec.~\ref{sec:SIS_lens}. 
In this case, $F(w,y)$ is given by Eq.~\eqref{eqn:amplification-factor-spherical-series},
and, again, it is only a function of $\lensMass$ and $y$.
To get the corresponding derivatives of $F(w, y)$, we need the derivative of $\Psi_n(w) = (-iw)^n/n!$ with respect to $w$, and the derivatives of $I_n(w,y)$ with respect to $w$ and $y$. 
The derivatives of the other terms in Eq.~\eqref{eqn:amplification-factor-spherical-series} are trivial.

The derivative of $\Psi_n(w) = (-iw)^n/n!$ with respect to $w$ is given by
\begin{equation}\label{eqn:derivative_Psi_n}
    \frac{\partial{\Psi_n(w)}}{\partial{w}} = \frac{(-i)^n}{(n-1)!}w^{n-1}\,.
\end{equation}

To obtain the derivative of $I_n(w,y)$ [Eq.~\eqref{eqn:important_integral}] with respect to $w$, we need to differentiate $_{1}F_{1}(N, 1;-iwy^2/2)$ with respect to $w$, where $N=(n+2)/2$. 
Using Eq.~\eqref{eqn:1F1_z}, we get
\begin{widetext}
  \begin{eqnarray}
    \label{eqn:derivative_importantIntegral_w}
        \frac{\partial I_n(w,y)}{\partial w} &=& \frac{N}{2}\left(\frac{2i}{w}\right)^N \Gamma(N) \left[ \left(-i\frac{y^2}{2}\right)\ _{1}F_{1}\left(N+1, 2; -i\frac{wy^2}{2}\right)
        -\frac{1}{w}\ _{1}F_{1}\left(N, 1; -i\frac{wy^2}{2}\right) \right]\,.\\
    \label{eqn:derivative_importantIntegral_y}
     \frac{\partial I_n(w,y)}{\partial y} &=& -iN\frac{wy}{2}\left(\frac{2i}{w}\right)^N \Gamma(N)\ _{1}F_{1}\left(N+1, 2; -i\frac{wy^2}{2}\right)\,.
    \end{eqnarray}
\end{widetext}

Using Eqs.~\eqref{eqn:derivative_Psi_n},~\eqref{eqn:derivative_importantIntegral_w}, and~\eqref{eqn:derivative_importantIntegral_y}, we get
\begin{widetext}
\begin{equation}\label{eqn:derivative_F_SIS_w}
        \frac{\partial F(w,y)}{\partial w} = \frac{\partial E(w,y)}{\partial w} \sum_{n=0}^{\infty} \Psi_n(w) I_n(w,y) +
        E(w,y) \sum_{n=0}^{\infty} \left[ \frac{\partial \Psi_n(w)}{\partial w} I_n(w,y) +  \Psi_n(w) \frac{\partial I_n(w,y)}{\partial w} \right]\,,
\end{equation}
\end{widetext}
where $E(w,y) = (w/i)\exp\left\{i w\left[y^2/2+\phi(y)\right]\right\}$. The partial derivative of $F(w,y)$ with respect to $\lensMass$ can be obtained by using the chain rule as before.

Finally, we have
\begin{multline}\label{eqn:derivative_F_SIS_y}
        \frac{\partial F(w,y)}{\partial y} = \frac{\partial E(w,y)}{\partial y} \sum_{n=0}^{\infty} \Psi_n(w) I_n(w,y)\\
        +E(w,y) \sum_{n=0}^{\infty} \Psi_n(w) \frac{\partial I_n(w,y)}{\partial y}\,.
\end{multline}

\bibliography{LISA_Lensing}

\end{document}